\documentclass[11pt,reqno,preprint]{article}
\usepackage{jheplike,epsfig,amssymb,amsmath,mathrsfs,hyperref,array,hhline}
\usepackage[makeroom]{cancel}
\usepackage{multirow,bbm}
\usepackage{feynmp-auto}
\usepackage{tikz-feynman}

\usepackage{mathtools}
\usepackage{caption}
\usepackage{subcaption}
\usepackage{placeins}

\makeatletter
\DeclareRobustCommand*{\bfseries}{
  \not@math@alphabet\bfseries\mathbf
  \fontseries\bfdefault\selectfont
  \boldmath
}

\def\eps{\epsilon}
\def\Li{\textrm{Li}}
\newcommand{\cEfe}[4]{{\mathcal{E}_4}\!\left(\begin{smallmatrix}#1\\#2\end{smallmatrix};#3,#4\right)}
 
\newcommand{\gamt}[3]{{\widetilde{\Gamma}}(\begin{smallmatrix}#1\\#2\end{smallmatrix};#3)} 
\newcommand{\fwboxL}[2]{\text{\makebox[#1][l]{$#2$}}}

\enlargethispage{2cm}
\renewcommand{\thefootnote}{\fnsymbol{footnote}}

\title{\vspace*{-1.3cm} \huge Functions Beyond Multiple Polylogarithms \\ for Precision Collider Physics \vspace{-.4cm}}

\author{\vspace{-.3cm} Jacob~L.~Bourjaily,$^{1,2}$ Johannes~Broedel,$^3$ Ekta~Chaubey,$^4$ Claude~Duhr,$^5$ Hjalte~Frellesvig,$^2$ Martijn~Hidding,$^6$ Robin~Marzucca,$^{2}$ Andrew~J.~McLeod,$^{7,8,}$\footnote[1]{a.mcleod@cern.ch} Marcus~Spradlin,$^{9,10}$ Lorenzo~Tancredi,$^{11}$ Cristian~Vergu,$^2$ Matthias~Volk,$^{12}$ Anastasia~Volovich,$^9$ Matt~von~Hippel,$^2$ Stefan~Weinzierl,$^{13}$ Matthias~Wilhelm,$^2$ \\ and Chi~Zhang$^2$}

\affiliation{$^1$ Institute for Gravitation and the Cosmos, Department of Physics, \\
Pennsylvania State University, University Park, PA 16802, USA}

\affiliation{$^2$ Niels Bohr International Academy, Niels Bohr Institute, \\
  Blegdamsvej 17, 2100 Copenhagen \O{}, Denmark}
  
\affiliation{$^3$ Institute for Theoretical Physics, ETH Zurich, Wolfgang-Pauli-Str.~27, 8093 Z\"urich, Switzerland}
  
\affiliation{$^4$ Physics Department, Torino University and INFN Torino, \\
Via Pietro Giuria 1, I-10125 Torino, Italy}

\affiliation{$^5$ Bethe Center for Theoretical Physics, Universit\"at Bonn, D-53115, Germany}

\affiliation{$^6$ Department of Physics and Astronomy, Uppsala University, SE-75120 Uppsala, Sweden}

\affiliation{$^7$ CERN, Theoretical Physics Department, 1211 Geneva 23, Switzerland}

\affiliation{$^8$ Mani L.~Bhaumik Institute for Theoretical Physics, \\
Department of Physics and Astronomy, UCLA, Los Angeles, California 90095, USA }

\affiliation{$^9$ Department of Physics, Brown University, Providence, RI 02912, USA}

\affiliation{$^{10}$ Brown Theoretical Physics Center, Brown University, Providence, RI 02912, USA}

\affiliation{$^{11}$ Physik Department, James-Franck-Stra{\ss}e 1, Technische Universit\"at M\"unchen, \\ D-85748 Garching, Germany}

\affiliation{$^{12}$ Institut f\"ur Physik, Humboldt-Universit\"at zu Berlin, \\ 
Zum Gro{\ss}en Windkanal 2, 12489 Berlin, Germany}

\affiliation{$^{13}$ PRISMA Cluster of Excellence, Institut f\"ur Physik, \\ Johannes Gutenberg-Universit\"at Mainz, D-55099 Mainz, Germany \vspace{-.5cm}}

\abstract{Feynman diagrams constitute one of the essential ingredients for making precision predictions for collider experiments. Yet, while the simplest Feynman diagrams can be evaluated in terms of multiple polylogarithms---whose properties as special functions are well understood---more complex diagrams often involve integrals over complicated algebraic manifolds. Such diagrams already contribute at NNLO to the self-energy of the electron, $t \bar{t}$ production, $\gamma \gamma$ production, and Higgs decay, and appear at two loops in the planar limit of maximally supersymmetric Yang-Mills theory. This makes the study of these more complicated types of integrals of phenomenological as well as conceptual importance.

In this white paper contribution to the Snowmass community planning exercise, we provide an overview of the state of research on Feynman diagrams that involve special functions beyond multiple polylogarithms, and highlight a number of research directions that constitute essential avenues for future investigation.  \vspace{-1.2cm} }

\preprint{\vspace*{-2cm}  \begin{flushright} BONN-TH-2022-05 $\vert$ UUITP-11/22 \\ CERN-TH-2022-029 $\vert$ TUM-HEP-1391/22 \\ HU-EP-22/08 $\vert$ MITP-22-022  
 \end{flushright}}

\begin{document}
\maketitle
\flushbottom
\begin{fmffile}{feyndiags}

\renewcommand*{\thefootnote}{\arabic{footnote}}

\section{Introduction}
\label{sec:introduction}

Over the last decade, it has become clear that a broad range of scattering amplitudes  can be expressed in terms of functions called multiple polylogarithms. This realization has led to enormous advances in our computational power, both in supersymmetric gauge theory~\cite{DelDuca:2009au,Caron-Huot:2011zgw,Dixon:2013eka,Dixon:2014voa,Golden:2014xqf,Dixon:2014iba,Dixon:2015iva,Henn:2016jdu,Caron-Huot:2016owq,Dixon:2016apl,Dixon:2016nkn,Bourjaily:2018aeq,Bourjaily:2019jrk,Bourjaily:2019vby,Caron-Huot:2018dsv,Drummond:2018caf,Abreu:2018aqd,Chicherin:2018yne,Henn:2019rgj,Caron-Huot:2019vjl,He:2019jee,McLeod:2020dxg,He:2020vob,He:2020lcu,Dixon:2020bbt,Chicherin:2021dyp,Golden:2021ggj,Li:2021bwg,Dixon:2021tdw} and in QCD, where substantial progress has recently been made computing $2 \to 3$ scattering processes to NNLO~\cite{Gehrmann:2015bfy,Ablinger:2017hst,Chicherin:2017dob,Gehrmann:2018yef,Badger:2018enw,Abreu:2018zmy,Abreu:2019odu,Badger:2019djh,Chawdhry:2019bji,Bonciani:2020tvf,Chicherin:2020oor,Abreu:2020cwb,Kallweit:2020gcp,Chawdhry:2020for,Abreu:2021oya,Badger:2021nhg,Agarwal:2021grm,Badger:2021owl,Chawdhry:2021mkw,Agarwal:2021vdh,Chawdhry:2021hkp,Gerlach:2021xtb,Czakon:2021mjy,Badger:2021imn,Badger:2021ega,Abreu:2021asb,Badger:2022ncb}, and obtaining the first results at N$^3$LO~\cite{Li:2014bfa,Ahmed:2014cla,Anastasiou:2015vya,Anastasiou:2016cez,Dreyer:2016oyx,Lee:2018nxa,Mistlberger:2018etf,Currie:2018fgr,Cieri:2018oms,Banerjee:2018lfq,Dulat:2018bfe,Dreyer:2018qbw,Gehrmann:2018odt,Mondini:2019gid,Duhr:2019kwi,Chen:2019lzz,Chen:2019fhs,Duhr:2020seh,Duhr:2020kzd,Duhr:2020sdp,Billis:2021ecs,Chen:2021isd,Camarda:2021ict,Duhr:2021vwj,Caola:2020dfu,Caola:2021rqz,Bargiela:2021wuy,Caola:2021izf} and beyond~\cite{Henn:2016wlm,Henn:2019rmi,Henn:2019swt,vonManteuffel:2020vjv,Lee:2021lkc,Lee:2022nhh}. In particular, much of this rapid progress has been facilitated by a deep understanding of the mathematical properties exhibited by multiple polylogarithms and the development of tools for working with these functions~\cite{Remiddi:1999ew,Gonch2,Goncharov:2010jf,Brown:2011ik,Duhr:2011zq,Duhr:2012fh,2011arXiv1101.4497D}, as well as by the availability of public codes for polylogarithmic integration and numerical evaluation~\cite{Bauer:2000cp,Gehrmann:2001pz,Gehrmann:2001jv,Vollinga:2004sn,Maitre:2005uu,Maitre:2007kp,Panzer:2014caa,Duhr:2019tlz}.  
 
However, it has long been known that special functions beyond multiple polylogarithms start to appear in scattering amplitudes at higher perturbative orders, especially in processes that depend on many kinematic variables. Namely, while multiple polylogarithms correspond to iterated integrals over rational functions, integrals over algebraic roots also start to arise in Feynman integrals at two loops.\footnote{While understanding the types of functions (such as hypergeometric functions) that appear in amplitudes nonperturbatively and in generic spacetime dimension $D$ also constitute interesting research avenues, we here focus on perturbative amplitudes in integer numbers of dimensions, or near integer numbers of dimensions in dimensional regularization.} These types of integrals appear even in the simplest quantum field theories, such as massless $\phi^4$ theory and maximally supersymmetric gauge theory, and correspondingly also appear in precision studies of the standard model, which will play an important role in searches for new physics in future collider programs~\cite{deBlas:2019rxi,EuropeanStrategyforParticlePhysicsPreparatoryGroup:2019qin}. The same types of integrals also appear in other contexts, such as string theory amplitudes. To make progress in all of these theories, we are thus led to consider classes of special functions that go beyond multiple polylogarithms.

The simplest integrals that force us out of the space of multiple polylogarithms involve square roots of polynomials that are cubic or quartic in one of the integration parameters, and thereby define an elliptic curve. A heroic amount of work has gone into understanding the first examples of Feynman integrals of this type~\cite{SABRY1962401,Broadhurst:1993mw,Berends:1993ee,Bauberger:1994by,Bauberger:1994hx,Bauberger:1994nk,Laporta:2004rb,Groote:2005ay,MullerStach:2012az,Groote:2012pa,Bloch:2013tra,Adams:2013kgc,Adams:2014vja,Adams:2015gva,Adams:2015ydq,Adams:2016xah,Remiddi:2016gno,Adams:2017tga,Adams:2017ejb,Bogner:2017vim,Remiddi:2017har,Bourjaily:2017bsb,Broedel:2017siw,Chen:2017soz,Adams:2018yfj,Broedel:2018iwv,Adams:2018bsn,Adams:2018kez,Honemann:2018mrb,Bogner:2019lfa,Broedel:2019hyg,Broedel:2019kmn}, and as a result these integrals are now under fairly good control. In particular, when only a single elliptic curve appears in a Feynman integral it is known how to express it in terms of elliptic generalizations of multiple polylogarithms, which can be formulated as iterated integrals over rational functions on the elliptic curve~\cite{brown2011multiple,Broedel:2017kkb} or as iterated integrals on a genus one Riemann surface~\cite{Bogner:2019lfa}. Moreover, much of the technology that has proven useful for working with multiple polylogarithms has been extended to elliptic polylogarithms, such as symbol calculus~\cite{2015arXiv151206410B,Broedel:2018qkq,Kristensson:2021ani} and tools for numerical evaluation~\cite{Hidding:2017jkk,Walden:2020odh,Hidding:2020ytt}. 

Integrals over more complicated algebraic quantities also appear in scattering amplitudes. These include square roots of higher-order polynomials of a single integration parameter~\cite{Huang:2013kh,Hauenstein:2014mda}, and square roots that depend on multiple integration parameters~\cite{Groote:2005ay,Brown:2010bw,Bloch:2014qca,Bloch:2016izu,mirrors_and_sunsets,Primo:2017ipr,Bourjaily:2018ycu,Bourjaily:2018yfy,Bourjaily:2019hmc}. In all known examples where algebraic roots involving multiple integration parameters occur, these roots have been found to describe Calabi-Yau manifolds. This has raised a number of important theoretical questions, such as whether integrals over higher-dimensional varieties that are not Calabi-Yau might also arise, and whether the geometry of these manifolds encode physical principles such as locality or causality in a systematic way. It has also raised the practical question of how best to evaluate these integrals, which occur as early as two loops even in massless theories~\cite{Bourjaily:2018yfy}.

In this white paper, we provide a broad review of the current state of knowledge about the types of integrals and special functions that appear beyond multiple polylogarithms in perturbative scattering amplitudes, and the technology that has been developed for dealing with them. We do this with the main goal of highlighting the many open questions and future research directions that deserve significant attention in the coming years. In particular, we emphasize the novel conceptual issues that are encountered when working with integrals over elliptic and Calabi-Yau manifolds, and the tradeoffs that are associated with different approaches to their evaluation. The importance of these questions is also highlighted, both for developing our conceptual understanding of perturbative quantum field theory and for increasing our capacity to make phenomenologically-relevant predictions for upcoming collider experiments. Just as the discovery that multiple polylogarithms provide the right language for certain classes of amplitudes led to a revolution in computational techniques, finding the right mathematical language to frame amplitudes beyond multiple polylogarithms can be expected to lead to equally important and exciting advances.

This paper is structured as follows. In section~\ref{sec:review_and_importance}, we give a brief introduction to how non-polylogarithmic integrals arise in scattering amplitudes, and in section~\ref{sec:integral_zoo} we catalog the best-studied examples of Feynman diagrams that give rise to such integrals. The current state of the art for dealing with elliptic polylogarithms and integrals over higher-dimensional varieties is then reviewed in section~\ref{sec:current_technology}. In section~\ref{sec:future_research}, we highlight a number of important open questions and directions for future research, and motivate their practical and theoretical importance. We end with a brief outlook in section~\ref{sec:conclusions}.

\section{Integrals Beyond Multiple Polylogarithms in Scattering Amplitudes}
\label{sec:review_and_importance}

We begin by illustrating how integrals that go beyond the space of multiple polylogarithms first arise in scattering amplitudes. To do so, let us recall that the integral expression corresponding to an $L$-loop Feynman diagram takes the form 
\begin{equation}
\mathcal{I} = \int \prod_{\ell} \frac{d^D k_\ell}{(2\pi)^D} \prod_{j} \frac{i}{[q_j(k,p)]^2 - m_j^2 + i \eps} \, , \label{eq:feynman_int_form}
\end{equation}
where $D$ is the spacetime dimension, $\ell$ indexes the $L$ loop momenta, and $j$ indexes the internal propagators. (We have assumed all particles are scalars.) The loop momentum associated with the $\ell^{\text{th}}$ loop is denoted by $k_\ell$, while $m_j$ and $q_j(k,p)$ respectively denote the mass and momentum flowing through the $j^{\text{th}}$ propagator. The specific form taken by each momentum $q_j(k,p)$ will depend on the topology of the Feynman diagram, but will in general involve only linear dependence on the external momenta (collectively denoted by $p$) and the loop momenta (collectively denoted by $k$). 

Feynman diagrams involving particles with nonzero spin will involve numerators that depend on the loop momenta. However, these integrals can always be reduced to scalar integrals using integration-by-parts (IBP) identities~\cite{Chetyrkin:1981qh,Laporta:2000dsw} and dimensional recurrence relations~\cite{Tarasov:1996br,Tarasov:1997kx}, at the cost of considering integrals in different spacetime dimensions and that involve propagators raised to higher (integer) powers. For this reason, we will only consider Feynman integrals with loop-momentum-independent numerators. 
 
We can attempt to evaluate an integral of the form~\eqref{eq:feynman_int_form} directly, using Feynman parameters. To translate to the Feynman parameter representation, we use the identity
\begin{equation}
\prod_{j=1}^E \frac{1}{A_j} = \int_{x_j \ge 0} \left[ d^{E-1} x_j \right] \frac{(E-1)!}{\big(\sum_{j=1}^E x_j A_j\big)^E} \, ,
\end{equation}
where $E$ denotes the total number of internal propagators, and $\left[ d^{E-1} x_j \right]$ is the canonical volume form on the projective space $\mathbb{RP}^{E-1}$ of Feynman parameters. After collecting all propagators into a single denominator factor using this identity, the integrals over loop momenta can be performed using standard techniques. This leaves what is sometimes referred to as the Symanzik form of the Feynman integral $\mathcal{I}$:
\begin{equation}
\mathcal{I} \sim \Gamma(E -  LD/2) \int_{x_j \ge 0} \left[ d^{E-1} x_j \right] \frac{\mathcal{U}^{E-(L+1)D/2}}{\mathcal{F}^{E-LD/2}} \, , \label{eq:symanzik}
\end{equation}  
where we have ignored an overall prefactor, and the $\mathcal{U}$ and $\mathcal{F}$ polynomials are defined by
\begin{align}
\mathcal{U} &= \sum_{\{ T \} \in \mathcal{T}_1 } \prod_{e_j \notin T} x_j \, , \\
\mathcal{F} &= \left[ \sum_{\{T_1,T_2\} \in \mathcal{T}_2} s_{T_1} \left( \prod_{e_j \notin T_1 \cup T_2} x_j \right) \right] + \mathcal{U} \sum_{e_j} x_j m_j^2 \, .
\end{align}
Here, $\mathcal{T}_k$ denotes the set of spanning $k$-forests of the Feynman diagram, and $s_{T_1}$ represents the squared sum of momentum flowing into the tree $T_1$. For more details on this representation, see for instance~\cite{Smirnov:2004ym}. 

One can always carry out at least one of the integrations in~\eqref{eq:symanzik}, since the $\mathcal{F}$ polynomial is at most quadratic in each integration variable (while the $\mathcal{U}$ polynomial is linear in each). In integer spacetime dimensions, this amounts to partial fractioning the integrand so that each factor in the denominator is linear in the chosen integration variable, after which the integral can be performed either rationally or logarithmically.\footnote{In dimensional regularization, one must first subtract possible divergences and expand in the dimensional regularization parameter.} In some cases, this strategy can be repeated and all integrals can be evaluated using the iterative definition of multiple polylogarithms:
\begin{equation} \label{eq:G_notation_1}
G_{a_1,\dots, a_w}(z) = \int_0^z \frac{dt}{t-a_1} G_{a_2,\dots, a_w}(t)\, ,
\end{equation}
where $G_{\emptyset}(z) = 1$, and the indices $a_i$ and the argument $z$ are complex numbers that can depend algebraically on external kinematics. When the first $w$ integrations involve $a_i$ that are all $0$, this definition diverges, and we instead define
\begin{equation} \label{eq:G_notation_2}
G_{\fwboxL{27pt}{{\underbrace{0,\dots,0}_{w}}}}(z) = \frac{\log^w z}{w!} \,. 
\end{equation}
For more details on multiple polylogarithms, see~\cite{Duhr:2014woa,Weinzierl:2022eaz}, and for a discussion of when all of the integrals in~\eqref{eq:symanzik} can be carried out in terms of multiple polylogarithms, see~\cite{Brown:2008um,Brown:2009ta,Panzer:2015ida,Bourjaily:2018aeq,Bourjaily:2021lnz}.

In more complicated Feynman integrals, partial fractioning will eventually lead to square roots in the denominator, which cannot be integrated using the definition~\eqref{eq:G_notation_1}. For instance, carrying out an integral of the form
\begin{align}
\int_0^\infty \frac{d x}{x^2+ 2 f x + g} &= \int_0^\infty \frac{dx}{2 \sqrt{f^2- g}}\left( \frac{1}{x + f - \sqrt{f^2- g}} - \frac{1}{x + f + \sqrt{f^2- g}} \right) \nonumber \\
&= \frac{1}{2 \sqrt{f^2- g}} \log \left( \frac{f+\sqrt{f^2- g}}{f-\sqrt{f^2- g}} \right) \, 
\end{align}
leads to a square root that depends on the coefficients $f$ and $g$. In situations where $f$ and $g$ depend on further integration variables, we end up with an integral over an algebraic variety that is parameterized by these integration variables. 

\begin{figure}
\begin{center}
    \begin{tikzpicture}
    \begin{feynman}
      \vertex (a) at (-1.5,0);
      \vertex (b) at ( 1.5,0);
      \vertex (e1) at (-3,0);
      \vertex (e2) at (3,0);
      \diagram* {
    (e1) -- [very thick, edge label=$p$] (a),
    (a) -- [half left, very thick, looseness=1.2, edge label=$m_1$] (b),
    (a) -- [very thick, edge label=$m_2$] (b),
    (a) -- [half right, very thick, looseness=1.2, edge label=$m_{3}$] (b),
    (b) -- [very thick] (e2),
   };
    \end{feynman}
    \end{tikzpicture}
\end{center}
\vspace{-.7cm}
\caption{The two-loop banana (or sunrise) diagram with generic internal masses.}
\label{fig:two_loop_banana}
\end{figure}
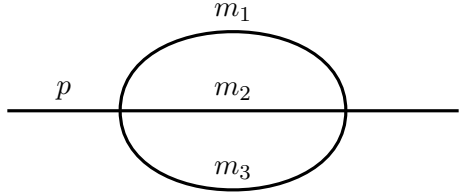

Consider, for instance, the two-loop banana (or sunrise) integral, shown in Figure~\ref{fig:two_loop_banana}. In two dimensions this integral is finite and the $\mathcal{U}$ polynomial drops out, leaving us with 
\begin{equation}
\mathcal{I}^{(2)}_{\text{ban}} = \int_{x_j \ge 0} \left[ d^2 x_j \right]  \frac{1}{x_1 x_2 x_3 \left( - p^2 + \left(\frac{1}{x_1} +  \frac{1}{x_2} +  \frac{1}{x_3}\right) \left(x_1 m_1^2 + x_2 m_2^2 + x_3 m_3^2 \right)\right)} \, , \label{eq:two_loop_banana}
\end{equation}
where $p$ is the momentum flowing into the diagram, and $m_j$ is the mass of the $j^{\text{th}}$ internal edge. Carrying out the integral over $x_3$, we find
\begin{equation}
\mathcal{I}^{(2)}_{\text{ban}} = \int_{x_i \ge 0} \Big[ d x_i \Big]  \frac{\log(X + y)-\log(X - y)}{y} \, , \label{eq:two_loop_banana_integrated}
\end{equation}
which depends on the polynomial
\begin{equation}
X =  m_1^2 x_1^2 + m_2^2 x_2^2 + x_1 x_2 (m_1^2 + m_2^2 + m_3^2 - p^2)  
\end{equation}
and a square root $y$, defined by the equation
\begin{equation}
y^2 = x_1^2 x_2^2 \left[ \left( -p^2 {+} m_3^2 {+} \left(\frac{1}{x_1}{+}\frac{1}{x_2} \right) \left(x_1 m_1^2 {+} x_2 m_2^2 \right)\right)^2\!\! - 4 m_3^2 \left(\frac{1}{x_1}{+}\frac{1}{x_2} \right) \left(x_1 m_1^2 {+} x_2 m_2^2 \right) \right] . \label{eq:two_loop_elliptic_curve}
\end{equation}
If we deprojectivize~\eqref{eq:two_loop_banana} by setting $x_2 = 1$, equation~\eqref{eq:two_loop_elliptic_curve} becomes easily recognizable as an elliptic curve in the variable $x_1$, with kinematic-dependent coefficients:
\begin{equation}
y^2 \big|_{x_2 =1} = m_1^4 x_1^4 + m_1^2 Y x_1^3 + Y^\prime x_1^2  + m_2^2 Y x_1 + m_2^4 \, ,
\end{equation}
where 
\begin{align}
Y &= 2 (m_1^2 + m_2^2 - m_3^2 - p^2) \, , \\
Y^\prime &= 2 m_1^2 m_2^2 -4 (m_1^2 + m_2^2) m_3^2 + (m_1^2 + m_2^2 + m_3^2 - p^2)^2 \, .
\end{align}
The remaining integral over $x_1$ thus takes us out of the space of multiple polylogarithms defined in~\eqref{eq:G_notation_1} and~\eqref{eq:G_notation_2}, and into the space of elliptic multiple polylogarithms. We defer a discussion of this space of functions to section~\ref{sec:elliptic_multiple_polylogs}.

It is also worth seeing how integrals over elliptic curves (and higher-dimensional manifolds) appear in the differential equations approach to computing Feynman integrals~\cite{KOTIKOV1991158,Kotikov:1993zf,Remiddi:1997ny,Gehrmann:1999as}. This approach, which is used in most modern calculations of Feynman integrals, first consists of expressing the set of integrals one needs for a given phenomenological process in terms of a minimal set of independent Feynman integrals, referred to as master integrals. This reduction is often done using IBP identities, as implemented in a number of publicly available computer codes~\cite{Smirnov:2008iw, Maierhofer:2017gsa, Anastasiou:2004vj, vonManteuffel:2012np, Lee:2012cn}, although other methods also exist (see for instance~\cite{Mastrolia:2018uzb, Frellesvig:2019uqt}). One then assembles the set of master integrals into a vector $f$ and takes the derivative of $f$ with respect to different kinematic variables (such as Mandelstam variables, or internal masses). Denoting these kinematic variables schematically by $x$, this will give rise to a system of differential equations that can be put in the form
\begin{align}
\frac{\partial}{\partial x} f = A f \, ,
\label{eq:difeq}
\end{align}
where the entries of the matrix $A$ are rational functions of the spacetime dimension $D$ and kinematic variables.\footnote{The kinematic derivatives of $f$ can always be put in the form~\eqref{eq:difeq}, since the action of these derivatives can be expressed in terms of Feynman integrals involving propagators raised to different powers, which can be reduced to the master integrals via IBPs. For a more pedagogical presentation of how differential equations of the form~\eqref{eq:difeq} can be generated, see for instance~\cite{Weinzierl:2022eaz}.} The matrix $A$ will be block triangular, since the Feynman integrals on the left can only couple to integrals on the right that involve a subset of the same propagators. 

While one can attempt to solve~\eqref{eq:difeq} using traditional differential equation methods, it has been found that there often exists a change of variables or way of redefining the basis of master integrals such that~\eqref{eq:difeq} is recast in the form
\begin{align}
\frac{\partial}{\partial x} f = \epsilon \mathcal{A} f \, ,
\label{eq:cannicaldifeq}
\end{align}
where $\epsilon$ is the small parameter in dimensional regularization, and where the matrix $\mathcal{A}$ has no further dependence on $\epsilon$. If all entries of $\mathcal{A}$ are $d\log$s of rational functions of the kinematic variables, this differential equation is considered to be in canonical form~\cite{Henn:2013pwa}, and can be directly integrated to polylogarithms.

In cases involving integrals over elliptic curves or higher-dimensional varieties, traditional methods for obtaining a canonical form~\cite{Henn:2014qga,Lee:2014ioa,Argeri:2014qva,Meyer:2016slj,Gituliar:2017vzm,Henn:2020lye,Chen:2020uyk} usually fail, or will lead to a matrix $\mathcal{A}$ that involves transcendental factors~\cite{Frellesvig:2021hkr}. For example, in the case of the two-loop banana integral, the $\epsilon$ dependence of the matrix $A$ can be factored out to realize a differential equation of the form~\eqref{eq:cannicaldifeq}, but this requires a non-algebraic change of variables that introduces periods of the elliptic curve~\eqref{eq:two_loop_elliptic_curve} into the matrix $\mathcal{A}$~\cite{Remiddi:2016gno,Adams:2016xah} (see~\cite{Weinzierl:2022eaz} for a pedagogical presentation of this example). In such cases, one can follow an alternate strategy by putting the system of differential equations into a form that is merely linear in $\epsilon$ (but that is no longer homogeneous). Then, one can iteratively construct the solution to these differential equations starting from the maximal cuts of each master integral, which provide solutions to the homogeneous part of these equations~\cite{Primo:2016ebd,Primo:2017ipr} and can be conveniently computed in the so-called Baikov representation~\cite{Baikov:1996iu,Frellesvig:2017aai,Harley:2017qut}. We defer further details about the types of functions these differential equations integrate into to section~\ref{sec:current_technology}. 

Before moving on, let us briefly highlight that elliptic polylogarithms (and integrals over higher-dimensional varieties) can sometimes be avoided even when algebraic roots appear during integration, by finding changes of variables that rationalize these roots. Even so, finding such a change of variables can prove hard in practice. We provide a discussion of when it is expected it can be done in section~\ref{sec:rationalizing_roots}. The algebraic roots that appear in Feynman integrals can also rationalize in special kinematic limits. For instance, in the $p^2 \to 0$ limit of the two-loop banana, the right side of~\eqref{eq:two_loop_elliptic_curve} becomes a perfect square and the elliptic curve degenerates:
\begin{equation}
y^2\big|_{p^2 \to 0} = \left( (x_1+x_2) (m_1^2 x_1 + m_2^2 x_2) - m_3^2 x_1 x_2 \right)^2 \, . \label{eq:two_loop_elliptic_curve_degeneration}
\end{equation}
In this limit, one can carry out the remaining integration in~\eqref{eq:two_loop_banana_integrated} in terms of multiple polylogarithms~\cite{Bloch:2013tra}.\footnote{The elliptic curve defined by~\eqref{eq:two_loop_elliptic_curve} also degenerates when $m_3 \to 0$. However, this limit is infrared divergent, so to compute it one should work in $D= 2-2\epsilon$ dimensions.}
While it can be useful to determine the kinematic limits in which these elliptic curves (or higher-dimensional varieties) degenerate, this can prove hard in practice when the corresponding square roots depend on many kinematic variables. Such questions thus constitute important directions for future work.


\section{The Non-Polylogarithmic Zoo}
\label{sec:integral_zoo}

Before describing the technology that has been developed for dealing with elliptic polylogarithms and integrals over higher-dimensional Calabi-Yau varieties, we briefly outline some of the best-studied examples of Feynman integrals that require this technology.
We focus on Feynman integrals in the context of particle physics calculations, although it is worth noting that elliptic integrals have also been observed in the classical limit of gravity~\cite{Bern:2021dqo,Bern:2021yeh}.

\subsection{The Sunrise and Banana Integrals}
\label{sec:banana-integrals}

\begin{figure}[t]
\begin{center}
    \begin{tikzpicture}
    \begin{feynman}
      \vertex (a) at (-1.5,0);
      \vertex (b) at ( 1.5,0);
      \vertex (e1) at (-3,0);
      \vertex (e2) at (3,0);
      \vertex (d1) at (0,0.4);
      \vertex (d2) at (0,.2);
      \vertex (d3) at (0,0);
      \draw[fill=black] (d1) circle(.2mm);
      \draw[fill=black] (d2) circle(.2mm);
      \draw[fill=black] (d3) circle(.2mm);
        \diagram* {
    (e1) -- [very thick, edge label=$p$] (a),
    (a) -- [half left, very thick, looseness=1.6, edge label=$m_1$] (b),
    (a) -- [half left, very thick, looseness=.8, edge label=$m_2$] (b),
    (a) -- [half right, very thick, looseness=.8, edge label=$m_{L-2}$] (b),
    (a) -- [half right, very thick, looseness=1.6, edge label=$m_{L-1}$] (b),
    (b) -- [very thick] (e2),
   };
    \end{feynman}
    \end{tikzpicture}
\end{center}
\vspace{-.7cm}
\caption{The $L$-loop banana integral, whose internal masses are all distinct, and which involves integrals over $(L{-}1)$-dimensional Calabi-Yau manifolds.}
\label{fig:banana_integrals}
\end{figure}
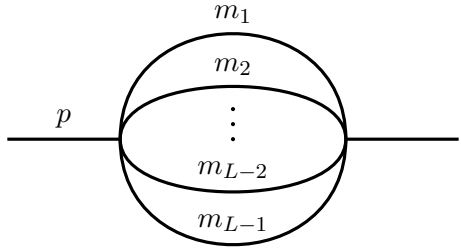

The class of integrals that have been studied in most depth are the banana (or sunrise) integrals with massive internal propagators, depicted in Figure~\ref{fig:banana_integrals}.
They are usually analyzed in $D=2-2 \epsilon$ dimensions, where they are simplest, as it is possible from these results to obtain the integral in $4-2 \epsilon$ dimensions using dimensional shift identities.
This infinite class of integrals has proven particularly useful as a nontrivial playground in which to explore the geometries and classes of special function that can appear in high-loop Feynman integrals.

As was discussed in section~\ref{sec:review_and_importance}, the two-loop banana integral already involves a family of elliptic curves that is parameterized by the internal and external masses~\cite{SABRY1962401,Broadhurst:1993mw,Caffo:1998du,Laporta:2004rb,Muller-Stach:2011qkg,Adams:2013nia,Remiddi:2013joa,Remiddi:2016gno}.
Early studies of this integral considered it as a generalized hypergeometric function, and derived various series and integral representations that allowed it to be evaluated numerically~\cite{Berends:1993ee,Broadhurst:1993mw,Bauberger:1994by,Bauberger:1994hx,Bauberger:1994nk}. More recently, the diagram involving equal internal masses has been evaluated in terms of elliptic dilogarithms~\cite{Laporta:2004rb,Bloch:2013tra} and iterated integrals over modular forms~\cite{Adams:2017ejb,Broedel:2018iwv}, while the diagram involving distinct masses has been evaluated in terms of integrals over the periods of the elliptic curve~\cite{Adams:2013kgc}, elliptic generalizations of multiple polylogarithms~\cite{Adams:2014vja,Adams:2015gva,Adams:2015ydq,Broedel:2017siw,Campert:2020yur}, and iterated integrals on the (properly compactified) moduli space of a Riemann surface of genus one with three marked points~\cite{Bogner:2019lfa}.\footnote{We will discuss these classes of special functions in section~\ref{sec:elliptic_multiple_polylogs}.} The family of elliptic curves that arises in this diagram has also been analyzed in great detail~\cite{Bloch:2016izu,mirrors_and_sunsets,Bourjaily:2018yfy,Frellesvig:2021vdl}.

At three loops, the banana integral with equal internal masses in strictly two dimensions also involves integrals over an elliptic curve. It has been evaluated in terms of elliptic polylogarithms and iterated integrals of modular forms~\cite{Broedel:2019kmn}. More generally, for distinct internal masses in dimensional regularization, the \(L\)-loop banana diagram involves integrals over a family of \((L{-}1)\)-dimensional Calabi-Yau manifolds, which are parameterized by external and internal masses~\cite{Bloch:2014qca,Bloch:2016izu,Bourjaily:2018yfy}.
A good understanding of the classes of special functions associated with integrals over these higher-dimensional varieties is currently lacking, and constitutes an important direction of ongoing research.
However, these Calabi-Yau geometries have already been used to derive systems of differential equations for the banana integrals~\cite{Klemm:2019dbm} and to express their solutions in terms of integrals over Calabi-Yau periods~\cite{Bonisch:2020qmm,Bonisch:2021yfw}.

\subsection{Traintracks}
\label{sec:traintrack}

\begin{figure}[t]
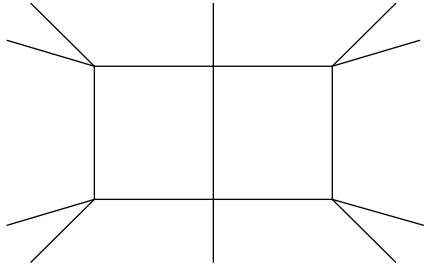

\begin{center}
\begin{fmfchar*}(300,140)
	\fmfset{dot_size}{0.18mm}
	 %
	\fmfforce{(.35w,.68h)}{p1}
	\fmfforce{(.5w,.68h)}{p2}
	\fmfforce{(.65w,.68h)}{p3}
	\fmfforce{(.65w,.32h)}{p4}
	\fmfforce{(.5w,.32h)}{p5}
	\fmfforce{(.35w,.32h)}{p6}
	%
	\fmfforce{(.24w,.75h)}{e1}
	\fmfforce{(.27w,.85h)}{e2}
	\fmfforce{(.5w,.85h)}{e3}
	\fmfforce{(.73w,.85h)}{e4}
	\fmfforce{(.76w,.75h)}{e5}
	\fmfforce{(.765w,.25h)}{e6}
	\fmfforce{(.73w,.15h)}{e7}
	\fmfforce{(.5w,.15h)}{e8}
	\fmfforce{(.27w,.15h)}{e9}
	\fmfforce{(.24w,.25h)}{e10}
	%
	\fmf{plain, width=.2mm}{p1,p2}
	\fmf{plain, width=.2mm}{p2,p3}
	\fmf{plain, width=.2mm}{p3,p4}
	\fmf{plain, width=.2mm}{p4,p5}
	\fmf{plain, width=.2mm}{p5,p6}
	\fmf{plain, width=.2mm}{p6,p1}
	%
	\fmf{plain, width=.2mm}{p2,p5}
	%
	\fmf{plain, width=.2mm}{p1,e1}
	\fmf{plain, width=.2mm}{p1,e2}
	\fmf{plain, width=.2mm}{p2,e3}
	\fmf{plain, width=.2mm}{p3,e4}
	\fmf{plain, width=.2mm}{p3,e5}
	\fmf{plain, width=.2mm}{p4,e6}
	\fmf{plain, width=.2mm}{p4,e7}
	\fmf{plain, width=.2mm}{p5,e8}
	\fmf{plain, width=.2mm}{p6,e9}
	\fmf{plain, width=.2mm}{p6,e10}
\end{fmfchar*}
\vspace{-1cm}
\end{center}
\caption{The massless ten-point double box integral, which involves integrals over an elliptic curve.}
\label{fig:elliptic_doublebox}
\end{figure}

The massless ten-point double box, shown in Figure~\ref{fig:elliptic_doublebox}, has also long been known to involve an elliptic curve from studies of its maximal cut and its representation in Mellin space~\cite{Paulos:2012nu,Caron-Huot:2012awx,Nandan:2013ip,Chicherin:2017bxc}. This Feynman diagram contributes to even the simplest quantum field theories, such as massless $\phi^4$ theory, maximally supersymmetric Yang-Mills theory, and integrable fishnet theories~\cite{Zamolodchikov:1980mb,Gurdogan:2015csr,Sieg:2016vap,Grabner:2017pgm}. In fact, it constitutes the leading contribution to one of the component amplitudes in planar maximally supersymmetric Yang-Mills theory, further highlighting its physical relevance~\cite{Caron-Huot:2012awx}. However, as this diagram depends on seven independent kinematic variables it has proven difficult to evaluate, and was only recently integrated in terms of elliptic polylogarithms~\cite{Kristensson:2021ani}. In the same work, the symbol of this integral was computed and found to have a remarkable similarity with the non-elliptic case. For instance, the symbol satisfies the same physical first entry condition~\cite{Gaiotto:2011dt} as multi-loop amplitudes in planar $\mathcal{N}=4$ supersymmetric Yang-Mills theory~\cite{Caron-Huot:2011zgw,Caron-Huot:2020bkp,He:2020vob,Li:2021bwg}, and elliptic symbol letters only appear in the last two entries; in particular, the last entry only involves elliptic integrals of the form $\int dx/y$.

\begin{figure}[t]
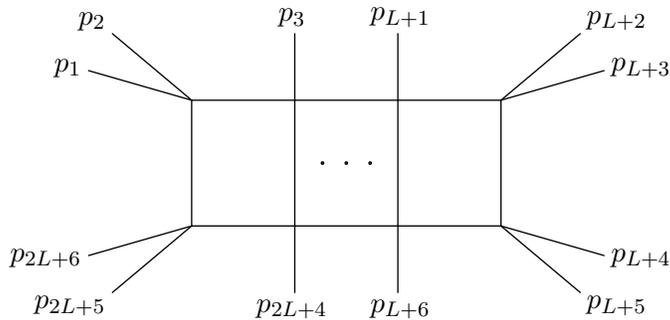

\begin{center}
\begin{fmfchar*}(300,140)
	\fmfset{dot_size}{0.18mm}
	 %
	\fmfforce{(.305w,.67h)}{p1}
	\fmfforce{(.435w,.67h)}{p2}
	\fmfforce{(.565w,.67h)}{p3}
	\fmfforce{(.695w,.67h)}{p4}
	\fmfforce{(.695w,.33h)}{p5}
	\fmfforce{(.565w,.33h)}{p6}
	\fmfforce{(.435w,.33h)}{p7}
	\fmfforce{(.305w,.33h)}{p8}
	%
	\fmfforce{(.47w,.5h)}{d1}
	\fmfforce{(.5w,.5h)}{d2}
	\fmfforce{(.53w,.5h)}{d3}
	%
	\fmfforce{(.175w,.75h)}{e1}
	\fmfforce{(.205w,.85h)}{e2}
	\fmfforce{(.435w,.85h)}{e3}
	\fmfforce{(.565w,.85h)}{e4}
	\fmfforce{(.795w,.85h)}{e5}
	\fmfforce{(.825w,.75h)}{e6}
	\fmfforce{(.825w,.25h)}{e7}
	\fmfforce{(.795w,.15h)}{e8}
	\fmfforce{(.565w,.15h)}{e9}
	\fmfforce{(.435w,.15h)}{e10}
	\fmfforce{(.205w,.15h)}{e11}
	\fmfforce{(.175w,.25h)}{e12}
	%
	\fmfforce{(.5w,.15h)}{l1}
	\fmfforce{(.2w,.5h)}{l3}
	\fmfforce{(.5w,.85h)}{l5}
	\fmfforce{(.8w,.5h)}{l7}
	%
	\fmf{plain, width=.2mm}{p1,p2}
	\fmf{plain, width=.2mm}{p2,p3}
	\fmf{plain, width=.2mm}{p3,p4}
	\fmf{plain, width=.2mm}{p4,p5}
	\fmf{plain, width=.2mm}{p5,p6}
	\fmf{plain, width=.2mm}{p6,p7}
	\fmf{plain, width=.2mm}{p7,p8}
	\fmf{plain, width=.2mm}{p8,p1}
	%
	\fmf{plain, width=.2mm}{p2,p7}
	\fmf{plain, width=.2mm}{p3,p6}
	%
	\fmf{plain, width=.2mm}{p1,e1}
	\fmf{plain, width=.2mm}{p1,e2}
	\fmf{plain, width=.2mm}{p2,e3}
	\fmf{plain, width=.2mm}{p3,e4}
	\fmf{plain, width=.2mm}{p4,e5}
	\fmf{plain, width=.2mm}{p4,e6}
	\fmf{plain, width=.2mm}{p5,e7}
	\fmf{plain, width=.2mm}{p5,e8}
	\fmf{plain, width=.2mm}{p6,e9}
	\fmf{plain, width=.2mm}{p7,e10}
	\fmf{plain, width=.2mm}{p8,e11}
	\fmf{plain, width=.2mm}{p8,e12}
	%
	\fmfdot{d1}
	\fmfdot{d2}
	\fmfdot{d3}
	%
	\fmfv{label={$p_1$}, label.dist=.1cm}{e1}
	\fmfv{label={$p_2$}, label.dist=.1cm}{e2}
	\fmfv{label={$p_3$}, label.dist=.1cm}{e3}
	\fmfv{label={$p_{L{+}1}$}, label.dist=.1cm}{e4}
	\fmfv{label={$p_{L{+}2}$}, label.dist=.1cm}{e5}
	\fmfv{label={$p_{L{+}3}$}, label.dist=.1cm}{e6}
	\fmfv{label={$p_{L{+}4}$}, label.dist=.1cm}{e7}
	\fmfv{label={$p_{L{+}5}$}, label.dist=.1cm}{e8}
	\fmfv{label={$p_{L{+}6}$}, label.dist=.1cm}{e9}
	\fmfv{label={$p_{2L{+}4}$}, label.dist=.1cm}{e10}
	\fmfv{label={$p_{2L{+}5}$}, label.dist=.1cm}{e11}
	\fmfv{label={$p_{2L{+}6}$}, label.dist=.1cm}{e12}
\end{fmfchar*}
\vspace{-.4cm}
\end{center}
\caption{The traintrack class of Feynman integrals, where all internal propagators are massless. The dots in the middle loop represent $2-L$ additional boxes with a single massless external lines emerging from each new vertex. At $L$ loops, this diagram involves integrals over an $(L{-}1)$-dimensional Calabi-Yau manifold.}
\label{fig:traintracks}
\end{figure}

The ten-point double box is just the first nontrivial member of the class of traintrack integrals, which occur at every loop order~\cite{Bourjaily:2018ycu}. These integrals are depicted in Figure~\ref{fig:traintracks}, where the dots in the middle loop represent $L{-} 2$ additional boxes, and each internal vertex is understood to involve a single massless external leg. Correspondingly, the number of kinematic parameters that appear in these integrals grows with the number of loops, making them harder to study beyond the first few loop orders. However, similar to the banana integrals, these Feynman diagrams are expected to involve integrals over increasingly high-dimensional manifolds as the loop order grows; in particular, the $L$-loop traintrack integral is conjectured to involve integrals over a Calabi-Yau manifold of dimension $L{-}1$~\cite{Bourjaily:2018ycu}. This expectation has been explored at low loop orders by computing residues until algebraic obstructions occur~\cite{Bourjaily:2018ycu}, and through an analysis of the leading singularity of the integral directly in momentum twistor space~\cite{Caron-Huot:2012awx,Vergu:2020uur}. In particular, the latter approach makes it clear that the leading singularity of the \(L\)-loop integral involves a family of Calabi-Yau \((L{-}1)\)-folds. As is the case for the two-loop double box, each of the traintrack diagrams constitutes the leading-order contribution to a component amplitude in planar $\mathcal{N}=4$ supersymmetric Yang-Mills theory~\cite{Bourjaily:2018ycu} (that is, no other diagrams of equal or lower loop order contribute to these component amplitudes), making them important candidates for further study.

\subsection{Tardigrades, Paramecia, and  Amoebas}
\label{sec:bestiary_integrals}

\begin{figure}[t]
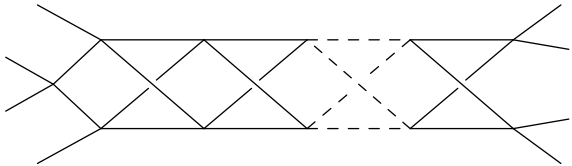

\begin{center}
\begin{fmfchar*}(300,60)
	\fmfset{dot_size}{0.18mm}
	 %
	\fmfforce{(.24w,.78h)}{p1}
	\fmfforce{(.37w,.78h)}{p2}
	\fmfforce{(.5w,.78h)}{p3}
	\fmfforce{(.63w,.78h)}{p4}
	\fmfforce{(.76w,.78h)}{p5}
	\fmfforce{(.76w,.22h)}{p6}
	\fmfforce{(.63w,.22h)}{p7}
	\fmfforce{(.5w,.22h)}{p8}
	\fmfforce{(.37w,.22h)}{p9}
	\fmfforce{(.24w,.22h)}{p10}
	\fmfforce{(.18w,.5h)}{p21}
	 %
	\fmfforce{(.30w,0.47385h)}{p11}
	\fmfforce{(.31w,0.52615h)}{p12}
	\fmfforce{(.43w,0.47385h)}{p13}
	\fmfforce{(.44w,0.52615h)}{p14}
	\fmfforce{(.56w,0.47385h)}{p15}
	\fmfforce{(.57w,0.52615h)}{p16}
	\fmfforce{(.69w,0.47385h)}{p17}
	\fmfforce{(.70w,0.52615h)}{p18}
	%
	\fmfforce{(.16w,1h)}{e1}
	\fmfforce{(.82w,1h)}{e2}
	\fmfforce{(.83w,.72h)}{e3}
	\fmfforce{(.83w,.28h)}{e4}
	\fmfforce{(.82w,.0h)}{e5}
	\fmfforce{(.12w,.33h)}{e6}
	\fmfforce{(.12w,.67h)}{e7}
	\fmfforce{(.16w,0h)}{e8}
	%
	\fmfforce{(.5w,.15h)}{l1}
	\fmfforce{(.2w,.5h)}{l3}
	\fmfforce{(.5w,.85h)}{l5}
	\fmfforce{(.8w,.5h)}{l7}
	%
	\fmf{plain, width=.2mm}{p1,p2}
	\fmf{plain, width=.2mm}{p2,p3}
	\fmf{dashes, width=.2mm}{p3,p4}
	\fmf{plain, width=.2mm}{p4,p5}
	\fmf{plain, width=.2mm}{p6,p7}
	\fmf{dashes, width=.2mm}{p8,p7}
	\fmf{plain, width=.2mm}{p8,p9}
	\fmf{plain, width=.2mm}{p9,p10}
	\fmf{plain, width=.2mm}{p10,p11}
	\fmf{plain, width=.2mm}{p2,p12}
	\fmf{plain, width=.2mm}{p9,p13}
	\fmf{plain, width=.2mm}{p3,p14}
	\fmf{dashes, width=.2mm}{p8,p15}
	\fmf{dashes, width=.2mm}{p4,p16}
	\fmf{plain, width=.2mm}{p7,p17}
	\fmf{plain, width=.2mm}{p5,p18}
	%
	\fmf{plain, width=.2mm}{p1,p9}
	\fmf{plain, width=.2mm}{p2,p8}
	\fmf{dashes, width=.2mm}{p7,p3}
	\fmf{plain, width=.2mm}{p4,p6}
	\fmf{plain, width=.2mm}{p1,p21}
	\fmf{plain, width=.2mm}{p10,p21}
	%
	\fmf{plain, width=.2mm}{p1,e1}
	\fmf{plain, width=.2mm}{p5,e2}
	\fmf{plain, width=.2mm}{p5,e3}
	\fmf{plain, width=.2mm}{p6,e4}
	\fmf{plain, width=.2mm}{p6,e5}
	\fmf{plain, width=.2mm}{p21,e6}
	\fmf{plain, width=.2mm}{p21,e7}
	\fmf{plain, width=.2mm}{p10,e8}
\end{fmfchar*}
\end{center}
\caption{The tardigrade diagrams, which exist for even numbers of loops and involve integrals over $(2L{-}2)$-dimensional Calabi-Yau manifolds at $L$ loops.}
\label{fig:tardigrades}
\end{figure}

\begin{figure}[t]
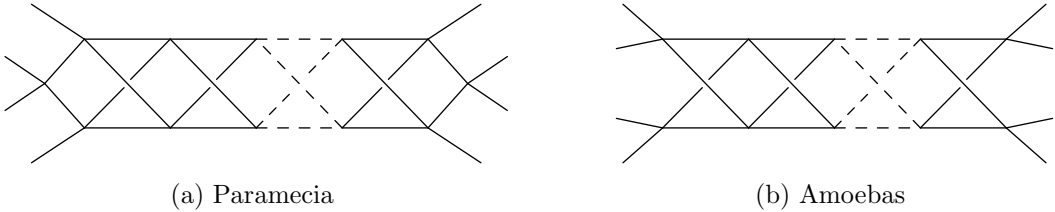

\hspace{-.5cm}
\subcaptionbox{Paramecia \label{fig:paramecia}}[.58\linewidth]
{\begin{fmfchar*}(250,60)
	 %
	\fmfforce{(.24w,.78h)}{p1}
	\fmfforce{(.37w,.78h)}{p2}
	\fmfforce{(.5w,.78h)}{p3}
	\fmfforce{(.63w,.78h)}{p4}
	\fmfforce{(.76w,.78h)}{p5}
	\fmfforce{(.76w,.22h)}{p6}
	\fmfforce{(.63w,.22h)}{p7}
	\fmfforce{(.5w,.22h)}{p8}
	\fmfforce{(.37w,.22h)}{p9}
	\fmfforce{(.24w,.22h)}{p10}
	\fmfforce{(.18w,.5h)}{p21}
	\fmfforce{(.82w,.5h)}{p22}
	 %
	\fmfforce{(.30w,0.47385h)}{p11}
	\fmfforce{(.31w,0.52615h)}{p12}
	\fmfforce{(.43w,0.47385h)}{p13}
	\fmfforce{(.44w,0.52615h)}{p14}
	\fmfforce{(.56w,0.47385h)}{p15}
	\fmfforce{(.57w,0.52615h)}{p16}
	\fmfforce{(.69w,0.47385h)}{p17}
	\fmfforce{(.70w,0.52615h)}{p18}
	%
	\fmfforce{(.16w,1h)}{e1}
	\fmfforce{(.84w,1h)}{e2}
	\fmfforce{(.88w,.67h)}{e3}
	\fmfforce{(.88w,.33h)}{e4}
	\fmfforce{(.84w,0h)}{e5}
	\fmfforce{(.12w,.33h)}{e6}
	\fmfforce{(.12w,.67h)}{e7}
	\fmfforce{(.16w,0h)}{e8}
	%
	\fmfforce{(.5w,.15h)}{l1}
	\fmfforce{(.2w,.5h)}{l3}
	\fmfforce{(.5w,.85h)}{l5}
	\fmfforce{(.8w,.5h)}{l7}
	%
	\fmf{plain, width=.2mm}{p1,p2}
	\fmf{plain, width=.2mm}{p2,p3}
	\fmf{dashes, width=.2mm}{p3,p4}
	\fmf{plain, width=.2mm}{p4,p5}
	\fmf{plain, width=.2mm}{p6,p7}
	\fmf{dashes, width=.2mm}{p8,p7}
	\fmf{plain, width=.2mm}{p8,p9}
	\fmf{plain, width=.2mm}{p9,p10}
	\fmf{plain, width=.2mm}{p10,p11}
	\fmf{plain, width=.2mm}{p2,p12}
	\fmf{plain, width=.2mm}{p9,p13}
	\fmf{plain, width=.2mm}{p3,p14}
	\fmf{dashes, width=.2mm}{p8,p15}
	\fmf{dashes, width=.2mm}{p4,p16}
	\fmf{plain, width=.2mm}{p7,p17}
	\fmf{plain, width=.2mm}{p5,p18}
	%
	\fmf{plain, width=.2mm}{p1,p9}
	\fmf{plain, width=.2mm}{p2,p8}
	\fmf{dashes, width=.2mm}{p7,p3}
	\fmf{plain, width=.2mm}{p4,p6}
	\fmf{plain, width=.2mm}{p1,p21}
	\fmf{plain, width=.2mm}{p10,p21}
	\fmf{plain, width=.2mm}{p5,p22}
	\fmf{plain, width=.2mm}{p6,p22}
	%
	\fmf{plain, width=.2mm}{p1,e1}
	\fmf{plain, width=.2mm}{p5,e2}
	\fmf{plain, width=.2mm}{p22,e3}
	\fmf{plain, width=.2mm}{p22,e4}
	\fmf{plain, width=.2mm}{p6,e5}
	\fmf{plain, width=.2mm}{p21,e6}
	\fmf{plain, width=.2mm}{p21,e7}
	\fmf{plain, width=.2mm}{p10,e8}
\end{fmfchar*}}
\hspace{-1.5cm}
\subcaptionbox{Amoebas \label{fig:amoebas}}[.58\linewidth]
{\begin{fmfchar*}(250,60)
	\fmfset{dot_size}{0.18mm}
	 %
	\fmfforce{(.24w,.78h)}{p1}
	\fmfforce{(.37w,.78h)}{p2}
	\fmfforce{(.5w,.78h)}{p3}
	\fmfforce{(.63w,.78h)}{p4}
	\fmfforce{(.76w,.78h)}{p5}
	\fmfforce{(.76w,.22h)}{p6}
	\fmfforce{(.63w,.22h)}{p7}
	\fmfforce{(.5w,.22h)}{p8}
	\fmfforce{(.37w,.22h)}{p9}
	\fmfforce{(.24w,.22h)}{p10}
	 %
	\fmfforce{(.30w,0.47385h)}{p11}
	\fmfforce{(.31w,0.52615h)}{p12}
	\fmfforce{(.43w,0.47385h)}{p13}
	\fmfforce{(.44w,0.52615h)}{p14}
	\fmfforce{(.56w,0.47385h)}{p15}
	\fmfforce{(.57w,0.52615h)}{p16}
	\fmfforce{(.69w,0.47385h)}{p17}
	\fmfforce{(.70w,0.52615h)}{p18}
	%
	\fmfforce{(.18w,1h)}{e1}
	\fmfforce{(.82w,1h)}{e2}
	\fmfforce{(.83w,.72h)}{e3}
	\fmfforce{(.83w,.28h)}{e4}
	\fmfforce{(.82w,.0h)}{e5}
	\fmfforce{(.18w,0h)}{e6}
	\fmfforce{(.17w,.28h)}{e7}
	\fmfforce{(.17w,.72h)}{e8}
	%
	\fmfforce{(.5w,.15h)}{l1}
	\fmfforce{(.2w,.5h)}{l3}
	\fmfforce{(.5w,.85h)}{l5}
	\fmfforce{(.8w,.5h)}{l7}
	%
	\fmf{plain, width=.2mm}{p1,p2}
	\fmf{plain, width=.2mm}{p2,p3}
	\fmf{dashes, width=.2mm}{p3,p4}
	\fmf{plain, width=.2mm}{p4,p5}
	\fmf{plain, width=.2mm}{p6,p7}
	\fmf{dashes, width=.2mm}{p8,p7}
	\fmf{plain, width=.2mm}{p8,p9}
	\fmf{plain, width=.2mm}{p9,p10}
	\fmf{plain, width=.2mm}{p10,p11}
	\fmf{plain, width=.2mm}{p2,p12}
	\fmf{plain, width=.2mm}{p9,p13}
	\fmf{plain, width=.2mm}{p3,p14}
	\fmf{dashes, width=.2mm}{p8,p15}
	\fmf{dashes, width=.2mm}{p4,p16}
	\fmf{plain, width=.2mm}{p7,p17}
	\fmf{plain, width=.2mm}{p5,p18}
	%
	\fmf{plain, width=.2mm}{p1,p9}
	\fmf{plain, width=.2mm}{p2,p8}
	\fmf{dashes, width=.2mm}{p7,p3}
	\fmf{plain, width=.2mm}{p4,p6}
	%
	\fmf{plain, width=.2mm}{p1,e1}
	\fmf{plain, width=.2mm}{p5,e2}
	\fmf{plain, width=.2mm}{p5,e3}
	\fmf{plain, width=.2mm}{p6,e4}
	\fmf{plain, width=.2mm}{p6,e5}
	\fmf{plain, width=.2mm}{p10,e6}
	\fmf{plain, width=.2mm}{p10,e7}
	\fmf{plain, width=.2mm}{p1,e8}
\end{fmfchar*}}
\caption{The paramecia and amoeba diagrams, which exist for odd numbers of loops and involve integrals over $(2L{-}2)$-dimensional Calabi-Yau manifolds at $L$ loops.}
\label{fig:paramecia_and_amoebas}
\end{figure}

In the nonplanar sector, three further classes of massless Feynman integrals have been shown to involve integrals over Calabi-Yau manifolds~\cite{Bourjaily:2018yfy}. They are known as the tardigrade, paramecium, and amoeba diagrams, and are shown in Figures~\ref{fig:tardigrades} and~\ref{fig:paramecia_and_amoebas}. While the tardigrades only exist at even loop orders, and the paramecia and amoebas only exist at odd loop orders, these diagrams all involve $2(L+1)$ propagators at $L$ loops, and thus the $\mathcal{U}$ polynomial in~\eqref{eq:symanzik} drops out in four dimensions. This simplification makes it possible to carry out the integral over any three of the $x_i$ variables explicitly. As observed in~\cite{Bourjaily:2018yfy}, these integrations give rise to a square root that encodes a Calabi-Yau of dimension $2(L{-}1)$ for every choice of three variables one chooses to integrate out.\footnote{The one exception seems to be the three-loop amoeba diagram; in this case, at least one further integration can be performed in terms of multiple polylogarithms.}

While none of these diagrams has yet been evaluated beyond one loop, it is worth noting that they all contribute to massless $\phi^4$ theory and QCD. This implies that integrals over K3 surfaces can already appear in these theories at two loops (in the guise of the two-loop tardigrade diagram). However, this contribution may still drop out of any given amplitudes in these theories; for instance, while the two-loop tardigrade integral appears as a contact term of many of the diagrams that appear in the local integral representation of the $n$-particle MHV amplitude in $\mathcal{N}=4$ supersymmetric Yang-Mills theory presented in~\cite{Bourjaily:2019iqr,Bourjaily:2019gqu}, all K3 contributions can be seen to drop out of the full expression.

\subsection{Further Integrals at Two and Three Loops}
\label{sec:two_three_loop_integrals}

\begin{figure}[t]
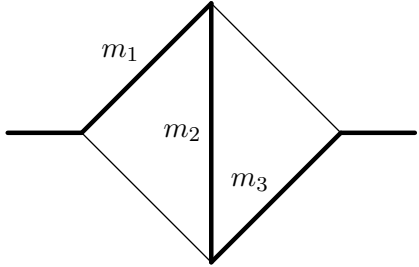

\begin{center}
\begin{fmfchar*}(140,140)

	\fmfforce{(.5w,.85h)}{v1}
	\fmfforce{(.85w,.5h)}{v2}
	\fmfforce{(.15w,.5h)}{v3}
	\fmfforce{(.5w,.15h)}{v4}
	
	\fmfforce{(-.05w,.5h)}{e1}
	\fmfforce{(1.05w,.5h)}{e2}
	
	\fmf{plain, width=.2mm}{v1,v2}
	\fmf{plain, width=.6mm,label={$m_1$},label.dist=.13cm}{v1,v3}
	\fmf{plain, width=.6mm,label={$m_3$},label.dist=.13cm}{v2,v4}
	\fmf{plain, width=.2mm}{v3,v4}
	\fmf{plain, width=.6mm,label={$m_2$},label.dist=.13cm}{v1,v4}
	\fmf{plain, width=.6mm}{v3,e1}
	\fmf{plain, width=.6mm}{v2,e2}

\end{fmfchar*} 
\vspace{-1cm}
\end{center}
\caption{The kite diagram with three distinct internal masses, which involves integrals over an elliptic curve. The three massive internal propagators are shown in bold, and the remaining two propagators are massless.}
\label{fig:kite}
\end{figure}

A number of additional integrals of phenomenological interest are known to involve elliptic curves at low loop orders. For instance, the kite diagram shown in Figure~\ref{fig:kite} contributes to the self-energy of the electron in QED when all three internal masses are equal ($m_1 = m_2 = m_3$). This diagram was first recognized to involve an elliptic integral nearly sixty years ago~\cite{SABRY1962401}, and has now been evaluated in terms of iterated integrals over products of elliptic integrals and polylogarithms~\cite{Remiddi:2016gno}, as iterated integrals over modular forms~\cite{Adams:2016xah,Adams:2017ejb,Bogner:2017vim,Bogner:2018uus,Adams:2018yfj}, and as elliptic multiple polylogarithms~\cite{Broedel:2018qkq}. Moreover, the full two-loop contribution to the self-energy of the electron has been evaluated in terms of iterated integrals of modular forms, which can be used to obtain $q$-expansions for efficient numerical evaluation~\cite{Honemann:2018mrb}. The kite integral with three distinct internal masses has also recently been computed in terms of elliptic multiple polylogarithms~\cite{Broedel:2019hyg}.

\begin{figure}[t]
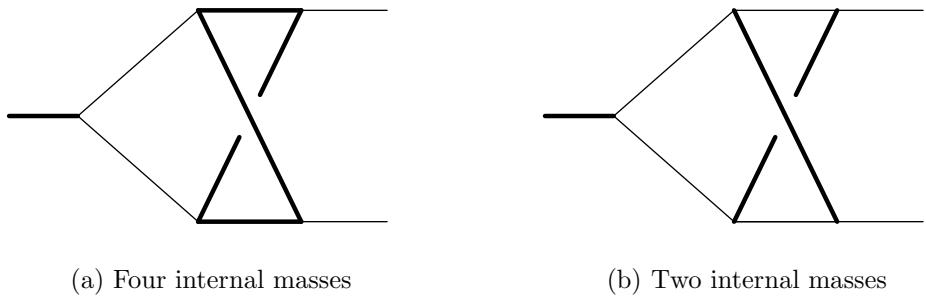

\begin{center}
\subcaptionbox{Four internal masses \label{fig:ttbar_loop}}[.45\linewidth]
{\begin{fmfchar*}(130,100)

	\fmfforce{(.1w,.5h)}{v1}
	\fmfforce{(.45w,.9h)}{v2}
	\fmfforce{(.75w,.9h)}{v3}
	\fmfforce{(.45w,.1h)}{v4}
	\fmfforce{(.75w,.1h)}{v5}
	\fmfforce{(.57w,.42h)}{v6}
	\fmfforce{(.63w,.58h)}{v7}
	
	\fmfforce{(-.1w,.5h)}{e1}
	\fmfforce{(1.w,.9h)}{e2}
	\fmfforce{(1.w,.1h)}{e3}
	
	\fmf{plain, width=.2mm}{v1,v2}
	\fmf{plain, width=.6mm}{v2,v3}
	\fmf{plain, width=.2mm}{v1,v4}
	\fmf{plain, width=.6mm}{v4,v5}
	\fmf{plain, width=.6mm}{v2,v5}
	\fmf{plain, width=.6mm}{v3,v7}
	\fmf{plain, width=.6mm}{v4,v6}

	\fmf{plain, width=.6mm}{e1,v1}
	\fmf{plain, width=.2mm}{e2,v3}
	\fmf{plain, width=.2mm}{e3,v4}

\end{fmfchar*}}
\subcaptionbox{Two internal masses \label{fig:electroweak_ff}}[.45\linewidth]
{\begin{fmfchar*}(130,100)

	\fmfforce{(.1w,.5h)}{v1}
	\fmfforce{(.45w,.9h)}{v2}
	\fmfforce{(.75w,.9h)}{v3}
	\fmfforce{(.45w,.1h)}{v4}
	\fmfforce{(.75w,.1h)}{v5}
	\fmfforce{(.57w,.42h)}{v6}
	\fmfforce{(.63w,.58h)}{v7}
	
	\fmfforce{(-.1w,.5h)}{e1}
	\fmfforce{(1.w,.9h)}{e2}
	\fmfforce{(1.w,.1h)}{e3}
	
	\fmf{plain, width=.2mm}{v1,v2}
	\fmf{plain, width=.2mm}{v2,v3}
	\fmf{plain, width=.2mm}{v1,v4}
	\fmf{plain, width=.2mm}{v4,v5}
	\fmf{plain, width=.6mm}{v2,v5}
	\fmf{plain, width=.6mm}{v3,v7}
	\fmf{plain, width=.6mm}{v4,v6}

	\fmf{plain, width=.6mm}{e1,v1}
	\fmf{plain, width=.2mm}{e2,v3}
	\fmf{plain, width=.2mm}{e3,v4}

\end{fmfchar*}}
\vspace{-.4cm}
\end{center}
\caption{Nonplanar two-loop integrals involving both massless and massive internal propagators (indicated by normal and bold lines, respectively). When all internal masses are the same, these diagrams can be expressed in terms of elliptic multiple polylogarithms.}
\label{fig:ttbar}
\end{figure}

The non-planar three-point integral shown in Figure~\ref{fig:ttbar_loop}, which has four integral propagators of the same mass, also involves an integral over an elliptic curve~\cite{vonManteuffel:2017hms,Mastrolia:2018uzb,Broedel:2019hyg}. This diagram contributes to $t\bar{t}$ production and $\gamma \gamma$ production in gluon fusion through a massive top-quark loop, and is thus relevant to our understanding of processes being investigated at the LHC. In recent years, it has been computed in terms of integrals over products of elliptic integrals and polylogarithms~\cite{vonManteuffel:2017hms} as well as in terms of elliptic multiple polylogarithms~\cite{Broedel:2019hyg}. The integral shown in Figure~\ref{fig:electroweak_ff}, which has only two massive internal propagators, has also been studied in the context of electroweak form factors~\cite{Aglietti:2004tq}, and was recently computed in terms of elliptic multiple polylogarithms~\cite{Broedel:2019hyg}. 

\begin{figure}[b]
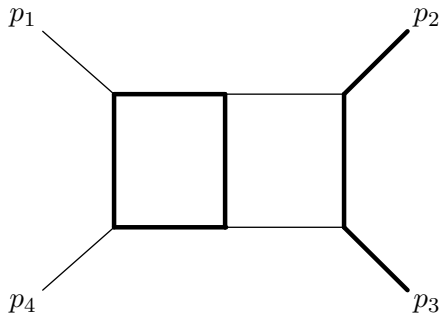

\begin{center}
\begin{fmfchar*}(300,140)
	 %
	\fmfforce{(.36w,.68h)}{p1}
	\fmfforce{(.5w,.68h)}{p2}
	\fmfforce{(.65w,.68h)}{p3}
	\fmfforce{(.65w,.32h)}{p4}
	\fmfforce{(.5w,.32h)}{p5}
	\fmfforce{(.5w,.32h)}{p6}
	\fmfforce{(.36w,.32h)}{p7}
	%
	\fmfforce{(.27w,.85h)}{e1}
	\fmfforce{(.73w,.85h)}{e2}
	\fmfforce{(.73w,.15h)}{e3}
	\fmfforce{(.27w,.15h)}{e4}
	%
	\fmfforce{(.5w,.15h)}{l1}
	\fmfforce{(.2w,.5h)}{l3}
	\fmfforce{(.5w,.85h)}{l5}
	\fmfforce{(.8w,.5h)}{l7}
	%
	\fmf{plain, width=.6mm}{p1,p2}
	\fmf{plain, width=.2mm}{p2,p3}
	\fmf{plain, width=.6mm}{p3,p4}
	\fmf{plain, width=.2mm}{p4,p5}
	\fmf{plain, width=.6mm}{p5,p6}
	\fmf{plain, width=.6mm}{p6,p7}
	\fmf{plain, width=.6mm}{p7,p1}
	%
	\fmf{plain, width=.6mm}{p2,p5}
	%
	\fmf{plain, width=.2mm}{p1,e1}
	\fmf{plain, width=.6mm}{p3,e2}
	\fmf{plain, width=.6mm}{p4,e3}
	\fmf{plain, width=.2mm}{p7,e4}
	%
	\fmfv{label={$p_1$}, label.dist=.1cm}{e1}
	\fmfv{label={$p_2$}, label.dist=.1cm}{e2}
	\fmfv{label={$p_3$}, label.dist=.1cm}{e3}
	\fmfv{label={$p_4$}, label.dist=.1cm}{e4}
\end{fmfchar*}
\end{center}
\vspace{-.5cm}
\caption{The double box integral that contributes to $t \bar t$ production via a top loop. All massive lines, shown in bold, are assigned the same mass. This integral involves integrals over three distinct elliptic curves.}
\label{fig:ttbar_doublebox}
\end{figure}

Versions of the double box integral with internal masses and various numbers of external legs have also been studied, both at the level of cuts~\cite{Caron-Huot:2012awx,Sogaard:2014jla} and iterated integrals~\cite{Bonciani:2016qxi,Adams:2018bsn,Adams:2018kez}. These diagrams are relevant to Higgs decay~\cite{Bonciani:2016qxi}, and to $t \bar{t}$ production via a top loop~\cite{Adams:2018bsn,Adams:2018kez}. Of particular interest is the integral studied in~\cite{Adams:2018bsn,Adams:2018kez}, which we have depicted in Figure~\ref{fig:ttbar_doublebox}. Unlike the other integrals we have encountered, this one involves three distinct elliptic curves, which can be associated with the subtopologies one gets by contracting all four horizontal propagators (giving rise to a two-loop banana topology), or the massless or massive pairs of horizontal propagators separately. While we expect that this integral cannot be expressed in terms of elliptic multiple polylogarithms, it has been computed to all orders in $\epsilon$ in terms of iterated integrals over kernels involving all three elliptic curves~\cite{Adams:2018bsn,Adams:2018kez}. Interestingly, when $(p_1+p_2)^2=m^2$, the elliptic curves all degenerate and this integral becomes expressible in terms of multiple polylogarithms, while in the $(p_1+p_4)^2 \to \infty$ limit all three elliptic curves become equal and the integral can be expressed as iterated integrals of modular forms. 

\begin{figure}[t]
\begin{center}
\subcaptionbox*{}[.4\linewidth]{
    \begin{tikzpicture}
    \begin{feynman}
      \vertex (a) at (-1.5,0);
      \vertex (b) at ( 1.5,0);
        \diagram* {
    (a) -- [half left, very thick, looseness=1.5, edge label=$m_1$] (b),
    (a) -- [half left, very thick, looseness=.55, edge label=$m_1$] (b),
    (a) -- [half right, very thick, looseness=.55, edge label=$m_1$] (b),
    (a) -- [half right, very thick, looseness=1.5, edge label=$m_2$] (b),
    };
    \end{feynman}
    \end{tikzpicture}
}
\subcaptionbox*{}[.4\linewidth]{ \vspace*{0.3cm}
    \begin{tikzpicture}
    \begin{feynman}
      \vertex (a) at (0,1.5);
      \vertex (b) at (1.5,0);
      \vertex (c) at (0,-1.5);
      \vertex (d) at (-1.5,0);
      \vertex (e) at (0,0);
      \vertex (f) at (1.29904,-.75);
      \vertex (g) at (-1.29904,-.75);
      \vertex (h) at (-1.6,1) {$m_1$};
      \vertex (i) at (1.6,1) {$m_2$};
      \vertex (j) at (.8,-.1) {$m_1$};
      \vertex (k) at (-.8,-.1) {$m_1$};
        \diagram* {
    (a) -- [quarter left, very thick] (b),
    (c) -- [quarter left] (d),
    (b) -- [quarter left] (c),
    (d) -- [quarter left, very thick] (a),
    (b) -- [quarter left, very thick, in=165.07, out=14.93, looseness=1] (f),
    (d) -- [quarter left, very thick, in=-165.07, out=-14.93, looseness=1] (g),
    (e) -- [] (a),
    (e) -- [very thick] (f),
    (e) -- [very thick] (g),
    };
    \end{feynman}
    \end{tikzpicture}
}
\end{center}
\vspace{-1.1cm}
\caption{Two of the diagrams that contribute to the $\rho$ parameter in the Standard Model and that involve integrals over an elliptic curve. }
\label{fig:rho_diagrams}
\end{figure}
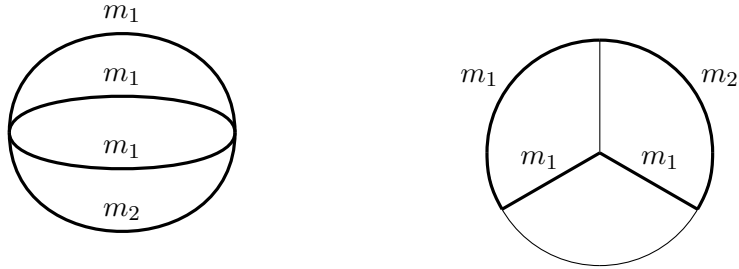

Integrals over elliptic curves also appear in perturbative contributions to the $\rho$ parameter, which encodes the difference between the vacuum polarizations of the W and Z bosons in the standard model~\cite{Grigo:2012ji,Ablinger:2017bjx,Blumlein:2018aeq}. These elliptic contributions first appear at three loops, when both the bottom and top quark masses are taken into account. Two of the diagrams in which these elliptic contributions appear are shown in Figure~\ref{fig:rho_diagrams}; these diagrams are vacuum graphs, and just depend on the quark masses $m_1$ and $m_2$. This perturbative correction to the $\rho$ parameter was first computed as a series expansion in the ratio of the two quark masses nearly a decade ago~\cite{Grigo:2012ji}, and so has long been under good numerical control. However, it has more recently been computed analytically in terms of certain `iterative non-iterative integrals'~\cite{Ablinger:2017bjx,Blumlein:2018aeq}, as well as in terms of elliptic multiple polylogarithms~\cite{Abreu:2019fgk}. Moreover, similar to the two-loop contribution to the self-energy of the electron, the three-loop correction to $\rho$ also turns out to be expressible in terms of the more restricted class of iterated integrals of modular forms~\cite{Abreu:2019fgk}.

\begin{figure}[t]
\begin{center}
    \begin{tikzpicture}
    \begin{feynman}
      \vertex (a) at (0,1.5);
      \vertex (b) at (1.5,0);
      \vertex (c) at (0,-1.5);
      \vertex (d) at (-1.5,0);
      \vertex (e) at (0,0);
      \vertex (f) at (1.29904,.75);
      \vertex (g) at (1.29904,-.75);
      \vertex (i) at (-1.29904,-.75);
      \vertex (j) at (-1.29904,.75);
      \vertex (e1) at (0,2.6);
      \vertex (e1a) at (-0.325866,2.5795);
      \vertex (e1b) at (0.325866,2.5795);
      \vertex (e2a) at (2.07098,1.57196);
      \vertex (e2b) at (2.39684,1.00754);
      \vertex (e3a) at (2.39684,-1.00754);
      \vertex (e3) at (2.25167,-1.3);
      \vertex (e3b) at (2.07098,-1.57196);
      \vertex (e4a) at (0.325866,-2.5795);
      \vertex (e4b) at (-0.325866,-2.5795);
      \vertex (e5) at (-2.25167,-1.3);
      \vertex (e5a) at (-2.07098,-1.57196);
      \vertex (e5b) at (-2.39684,-1.00754);
      \vertex (e6a) at (-2.39684,1.00754);
      \vertex (e6b) at (-2.07098,1.57196);
        \diagram* {
    (a) -- [quarter left] (b),
    (c) -- [quarter left] (d),
    (b) -- [quarter left] (c),
    (d) -- [quarter left] (a),
    (e) -- [] (a),
    (e) -- [] (g),
    (e) -- [] (i),
    (a) -- [] (e1),
    (f) -- [] (e2a),
    (f) -- [] (e2b),
    (g) -- [] (e3),
    (c) -- [] (e4a),
    (c) -- [] (e4b),
    (i) -- [] (e5),
    (j) -- [] (e6a),
    (j) -- [] (e6b),
    };
    \end{feynman}
    \end{tikzpicture}
\end{center}
\caption{The three-loop wheel diagram, which involves an integral over a Calabi-Yau threefold.}
\label{fig:wheel_diagram}
\end{figure}
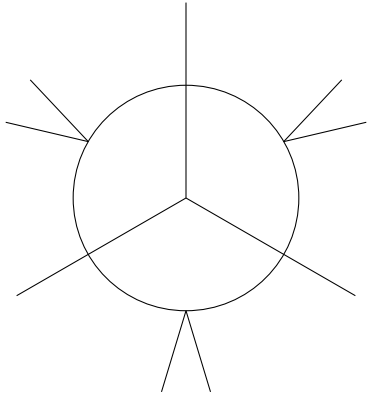

Finally, the massless three-loop wheel diagram shown in Figure~\ref{fig:wheel_diagram} has been shown to involve an integral over a Calabi-Yau threefold~\cite{Bourjaily:2018yfy,Bourjaily:2019hmc}. Like the classes of integrals described in sections~\ref{sec:traintrack} and~\ref{sec:bestiary_integrals}, this diagram contributes to massless $\phi^4$ theory and supersymmetric Yang-Mills theory. However, it has currently only been evaluated in special kinematic limits in which it can be expressed in terms of multiple polylogarithms~\cite{Bourjaily:2019hmc}. 

\subsection{String Amplitudes}

While integrals over nontrivial algebraic varieties appear in Feynman integrals due to the appearance of square root factors in the denominator, they also arise naturally in string theory as Green's functions.\footnote{In fact, it was in the context of string amplitudes that elliptic multiple polylogarithms first appeared in the physics literature~\cite{Broedel:2014vla}.} Namely, string amplitudes admit an expansion as integrals over surfaces of different genus, where the Green's functions on these surfaces determine the classes of differential forms that are integrated over. This makes string theory a perfect laboratory for exploring the classes of special functions associated with a particular surface, such as a complex torus or surfaces with higher genus.  

The simplest examples are tree-level open-string amplitudes \cite{Mafra:2011nv,Mafra:2011nw}, which are calculated by integrating over the positions of marked points (that give the locations of operator insertions) on the boundary of a disk. These integrals each take the form $\smash{\frac{dx}{x-a}}$, corresponding to the derivative of the appropriate Green's function. Accordingly, all tree-level open-string scattering amplitudes can be evaluated as multiple polylogarithms, using the definition from equation~\eqref{eq:G_notation_1}~\cite{Broedel:2013tta,Broedel:2013aza,Mafra:2016mcc}. Multiple zeta values also make an appearance in these amplitudes, as they arise as special values of multiple polylogarithms~\cite{Blumlein:2009cf,Brown:2011ik,Schlotterer:2012ny}. Using the methods described in~\cite{Broedel:2013aza,Mafra:2016mcc}, tree-level open-string amplitudes can be calculated at any multiplicity and to any order in the expansion in the inverse string tension $\alpha'$. 

Tree-level closed-string amplitudes are similar to tree-level open-string amplitudes, except that the marked points are now integrated over the whole Riemann sphere. Accordingly, the differential forms that appears in these amplitudes take the form $\frac{dz^2}{|z-a|^2}$, which leads to a subclass of multiple polylogarithms: the so-called single-valued multiple polylogarithms~\cite{BROWN2004527,BrownSVMPLs,brown2013singlevalued,2015arXiv151206410B}, which correspond to special linear combinations of multiple polylogarithms in which all branch cuts cancel.\footnote{Single-valued multiple polylogarithms also make an appearance in the multi-Regge limit of ${\cal N}=4$ supersymmetric Yang-Mills theory; see for instance \cite{Dixon:2012yy,Pennington:2012zj,Papathanasiou:2013uoa,Papathanasiou:2014yva,Broedel:2015nfp,DelDuca:2016lad,DelDuca:2018hrv,DelDuca:2018raq,DelDuca:2019tur}.} As a result, only single-valued multiple zeta values (which correspond to the sub-class of multiple zeta values that appear as special values of single-valued multiple polylogarithms) appear in these amplitudes~\cite{brown2013singlevalued,Stieberger:2013wea}. Using the so-called KLT relations \cite{Kawai:1985xq} and related ideas, tree-level closed-string amplitudes can in principle be calculated at any multiplicity and any order in $\alpha'$ from their open-string counterparts. 

In one-loop open-string amplitudes, one encounters integrals over the positions of marked points on the two boundaries of an annulus. These integrals must respect the double periodicity of this surface, while also remaining invariant under smooth deformations of the integration contour. This double requirement gives rise to an infinite tower of differential forms, which arise as the coefficients of the Eisenstein-Kronecker series (see equation~\eqref{eq:gamt_kernel} in the next section). Iterated integrals over these differential forms give rise to the same class of multiple elliptic polylogarithms encountered in Feynman integrals~\cite{Broedel:2014vla,Broedel:2017kkb}. These string amplitudes also involve elliptic multiple zeta values, which (despite their name) are functions of the modulus of the torus~\cite{Broedel:2015hia,Matthes:2016}. Using the recursive methods developed in~\cite{Mafra:2019xms,Broedel:2019gba}, one-loop open-string amplitudes can also be calculated algebraically at any multiplicity and to any order in $\alpha'$.  

One-loop closed-string amplitudes involve integrals over marked points on a torus, and correspondingly also give rise to elliptic multiple polylogarithms.\footnote{The points on an elliptic curve can be mapped to points on a complex torus, and so elliptic multiple polylogarithms can be equivalently formulated as iterated integrals on a torus. We review this correspondence in section~\ref{sec:elliptic_multiple_polylogs}.} As with tree-level closed-string amplitudes, they only seem to involve a restricted class of single-valued analogues of elliptic multiple polylogarithms~\cite{2014arXiv1407.5167B}. However, a clear understanding of the relation between the full space of elliptic multiple polylogarithms and their single-valued counterparts remains a topic of ongoing investigation. Far easier to access are the single-valued elliptic multiple zeta values, which make an appearance as so-called modular graph functions~\cite{DHoker:2015wxz}, which are also subjects of ongoing investigation (see for example \cite{Zerbini:2018hgs,Dorigoni:2019yoq,DHoker:2020hlp}). 

Going to higher loops, one encounters integrals over surfaces of higher genus. While the mathematics for dealing with these surfaces has in principle been developed and will involve special functions beyond elliptic multiple polylogarithms, few explicit results exist at two and three loops (but see for instance~\cite{DHoker:2001kkt,Gomez:2013sla}).


\section{Technological State of the Art}
\label{sec:current_technology}

A considerable amount of technology has already been developed for computing the types of Feynman integrals that were highlighted in the last section. This technology includes algorithmic tools for working with the types of special functions these integrals evaluate to analytically, as well as methods for their numerical evaluation. Given that the complicated manifolds that appear in these integrals arise as square roots, techniques for rationalizing such roots (and for knowing when they can be rationalized) also prove to be important. We begin by outlining what is known about the latter topic, before discussing the technology that exists for evaluating Feynman integrals involving square roots that cannot be rationalized.

\subsection{Rationalizing Square Roots}
\label{sec:rationalizing_roots}

An important question that arises in the computation of Feynman integrals is when square roots that appear in the course of integration can be rationalized by a change of variables. To address this question, it is useful to reframe the problem geometrically. To this end, consider a root $\sqrt{P(x_1,\dots, x_j)/Q(x_1,\dots, x_j)}$, where $P(x_1,\dots, x_j)$ and $Q(x_1,\dots, x_j)$ are polynomials in $j$ integration parameters, whose coefficients can depend on external kinematics. If we introduce an auxiliary variable $y$ that satisfies the relation
\begin{equation}
 h(x_1,\dots,x_j,y) = y^2 Q(x_1,\dots, x_j) - P(x_1,\dots, x_j) = 0 \, ,
\end{equation}
we define an affine hypersurface in $\mathbb{C}^{j+1}$. Rationalizing $y$ amounts to finding a rational parameterization of this hypersurface.

It is usually convenient to consider this problem in projective space, which can be done by introducing a variable $x_0$ and defining 
\begin{align}
\tilde{h}(x_0,x_1,\dots,x_j,y) &= x_0^d\, h(x_1/x_0,\dots,x_j/x_0,y/x_0) \\
&= y^2 \tilde{Q}(x_0,x_1,\dots, x_j) - \tilde{P}(x_0,x_1,\dots, x_j) \, ,
\end{align}
where $d$ is the overall power of the highest-degree monomial in $h(x_1,\dots,x_j,y)$, and we have implicitly defined a new pair of polynomials $\tilde{P}(x_0,x_1,\dots, x_j)$ and $\tilde{Q}(x_0,x_1,\dots, x_j)$. The homogeneous polynomial $\tilde{h}(x_0,x_1,\dots,x_j,y)$ defines a hypersurface $\smash{V(\tilde{h})}$ in $\mathbb{P}^{j+1}$, namely 
\begin{equation}
V(\tilde{h}) = \left\{ [x_0,x_1,\dots,x_j,y] \in \mathbb{P}^{j+1} \, \Big|\, \tilde{h}(x_0,x_1,\dots,x_j,y) = 0 \right\}   \, ,
\end{equation}
where $[x_0,x_1,\dots,x_j,y]$ denotes a point in homogeneous coordinates. Finding a rational parameterization of this surface then requires finding a rational map
\begin{equation}
\phi: \mathbb{P}^j \to V(\tilde{h}) , \qquad t \mapsto [\phi_0(t),\dots,\phi_{j+1}(t)]
\end{equation}
such that
\begin{equation}
\tilde{h}(\phi_0(t),\dots,\phi_{j+1}(t)) = 0 \quad \text{for all} \quad t \in \mathbb{P}^{j} \, .
\end{equation}
Given such a map, the ratio $\tilde{P}(\phi_0(t),\dots, \phi_{j}(t))/\tilde{Q}(\phi_0(t),\dots, \phi_{j}(t))$ must evaluate to a perfect square, since it is equal to $\phi_{j+1}(t)^2$ and we have required $\phi_{j+1}(t)$ to be rational. The map $\phi$ also rationalizes $y$ in the original affine space, via the change of variables $x_i  = \phi_i(t)/\phi_0(t)$. 

Let's see how this works in a simple example. If we encounter a root taking the form $\smash{\sqrt{a x_1^2 + b x_1 + c}}$, we can go through the above steps to find that it encodes a projective hypersurface in $\mathbb{P}^2$ defined by
\begin{equation}
\tilde{h}(x_0,x_1,y) = y^2  - a x_1^2 - b x_1 x_0 - c x_0^2  = 0 \, .
\end{equation}
The map $\phi$ can then be chosen to be
\begin{align}
\phi_0([t_0,t_1]) &= a t_0^2 - t_1^2 \, , \\
\phi_1([t_0,t_1]) &= a \sqrt{c} \, t_0^2 - t_1 (b t_0 - \sqrt{c} \, t_1) \, , \\
\phi_2([t_0,t_1]) &= -t_0 (b t_0 - 2 \sqrt{c} \, t_1) \, ,
\end{align} 
where $[t_0,t_1]$ are homogeneous coordinates in $\mathbb{P}^1$. It is not hard to check that setting $x = \phi_1([t_0,t_1])/\phi_0([t_0,t_1])$ rationalizes the original root, as desired. 

For most square roots, no rational parameterization exists. Consider for instance the class of roots with $j=1$, which define plane curves. In these cases, $V(\tilde{h})$ admits a rational parameterization if and only if the genus of this curve is zero~\cite{Clebsch1865}, where the genus of a curve of degree $d$ with only ordinary singularities can be computed as
\begin{equation}
g = \frac{(d - 1)(d - 2)}{2} - \sum  \frac{r_p (r_p - 1)}{2} \, .
\end{equation}
The sum in this formula is over all singular points $p$ in $V(\tilde{h})$, and $r_p$ denotes the multiplicity of the point $p$ (namely, the number of tangents to $V(\tilde{h})$ at $p$). It follows that a square root encoding a curve with only ordinary points can be rationalized if and only if
\begin{equation}
\sum  r_p (r_p - 1) = (d - 1)(d - 2) \, .
\end{equation}
Algorithms exist for finding rational parameterizations of curves that satisfy this criterion, and are implemented in the {\sc Singular} library {\tt paraplanecurves.lib}~\cite{DGPS,paraplanecurves}.
 
For $j=2$, a rational parameterization of $V(\tilde{h})$ again exists if and only if certain invariants of $V(\tilde{h})$ vanish~\cite{Castelnuovo} (see also~\cite{Besier:2020hjf}). An algorithm for computing these invariants and (when it exists) a rational parameterization is given in~\cite{Schicho:1998}. For $j > 2$, fewer results are known. However, when the hypersurface $V(\tilde{h})$ is of degree $d$ and has a singularity of multiplicity $d{-}1$, a rational parameterization can be computed using the algorithm presented in~\cite{Besier:2018jen} and implemented in the package {\sc RationalizeRoots}~\cite{Besier:2019kco}. Notably, while the computation of Feynman integrals often requires rationalizing multiple roots (see for instance~\cite{Becchetti:2017abb,Besier:2019hqd,Heller:2019gkq,Bourjaily:2019igt}), this algorithm can be applied to these roots iteratively.

\subsection{Iterated Integrals Involving Elliptic Curves}
\label{sec:elliptic_multiple_polylogs}

Of course, many of the algebraic roots that appear in the integration of Feynman integrals cannot be rationalized. The first important class of square roots for which this is true are those encoding non-degenerate elliptic curves. To evaluate Feynman integrals that involve such roots, we are thus led to consider iterated integrals over these elliptic curves.

Before introducing the elliptic generalizations of multiple polylogarithms, let us define some of the quantities that characterize elliptic curves, which correspond to the zero-locus of a polynomial equation $y^{2}=P_{n}(x)$ of degree $n=3$ or $4$. We focus on the $n=4$ case, where we have 
\begin{equation} \label{eq:elliptic_curve_ai}
y^2 = (x-a_1)(x-a_2)(x-a_3)(x-a_4) \, 
\end{equation}
with four distinct branch points $a_i$.\footnote{Note that, for $y^2$ to be real, these branch points must be in one of a few configurations in the complex plane. For more details, see~\cite{Broedel:2019hyg}.}  
The $n=3$ case can be obtained by sending one of the branch points, say $a_{4}$, to infinity. We define the periods of this elliptic curve to be
\begin{align}
\omega_1 &= 2 c_4 \int_{a_2}^{a_3} \frac{dx}{y} = 2 K (\lambda)\, \label{eq:period_1} \, ,\\
\omega_2 &= 2 c_4 \int_{a_1}^{a_2} \frac{dx}{y} = 2 i K(1- \lambda) \, , \label{eq:period_2}
\end{align}
where 
\begin{align}
\lambda = \frac{a_{14} a_{23}}{a_{13} a_{24}} \, , \qquad c_4 = \frac{1}{2} \sqrt{a_{13} a_{24}} \, , \qquad a_{ij} = a_i - a_j \, ,
\end{align}
and 
\begin{align}
K (\lambda) =  \int_0^1 \frac{dt}{\sqrt{(1-t^2)(1-\lambda t^2)}}\, 
\end{align}
denotes a complete elliptic integral of the first kind. 

Through the periods $\omega_1$ and $\omega_2$, the elliptic curve~\eqref{eq:elliptic_curve_ai} is naturally related to the torus one gets by quotienting the complex plane by the lattice $\Lambda = \mathbb{Z} \omega_1 + \mathbb{Z} \omega_2$. In practice, it is often convenient to normalize one of the periods to one, which leaves us with just a single nontrivial parameter, the modular parameter $\tau =\omega_{2}/\omega_{1}$. Then, the function
\begin{equation} \label{eq:elliptic_curve_torus_map}
z = \frac{c_{4}}{\omega_{1}} \int_{a_1}^x \frac{dx^\prime}{y}
\end{equation}
defines an isomorphism from points $(x,y)$ on the elliptic curve to points $z$ on the normalized torus $\mathbb{C}/(\mathbb{Z}+\mathbb{Z}\tau)$. As the lattice $\Lambda$ is invariant under modular transformations, the same elliptic curve will also be described by any pair of periods $\omega_1^\prime$ and $\omega_2^\prime$ that can be written as
\begin{align} \label{eq:modular}
\begin{pmatrix} \omega_2^\prime \\ \omega_1^\prime \end{pmatrix} = \begin{pmatrix} a & b \\ c & d \end{pmatrix} \begin{pmatrix} \omega_2 \\ \omega_1 \end{pmatrix}
\end{align}
for integer $a$, $b$, $c$, and $d$ such that $ad - bc = 1$. For a more detailed introduction to the properties of elliptic curves in the context of Feynman integrals, see~\cite{Weinzierl:2022eaz}.

Various classes of special functions have been defined in order to evaluate the integrals over elliptic curves that arise in Feynman integrals. One natural way of doing this is to generalize the definition of multiple polylogarithms in~\eqref{eq:G_notation_1} to iterated integrals that also depend on $y$~\cite{BeilinsonLevin,LevinRacinet2007,brown2011multiple,Broedel:2014vla,Matthes:2016,Broedel:2017kkb,Broedel:2017siw,Broedel:2018qkq}. One such class of elliptic multiple polylogarithms was introduced in~\cite{Broedel:2018qkq}, using the notation 
\begin{align}
\label{eq:cE4_def}
\cEfe{n_1 & \ldots & n_k}{b_1 & \ldots& b_k}{x}{\vec{a}} = \int_0^xdt\,\Psi_{n_1}(b_1,t,\vec a)\,\cEfe{n_2 & \ldots & n_k}{b_2 & \ldots& b_k}{t}{\vec a}\, ,
\end{align}
where $\cEfe{}{}{x}{\vec{a}} = 1$, the argument $\vec{a}$ represents the dependence on the four roots $a_i$ in~\eqref{eq:elliptic_curve_ai}, the indices $n_i$ are integers, and the arguments $x$ and $b_i$ are complex numbers that can depend on external kinematics. The allowed set of integration kernels $\Psi_{n}(b_1,t,\vec a)$ include those seen in the space of multiple polylogarithms,
\begin{align} \label{eq:mpl_kernel}
\Psi_{1}(b,t,\vec a) = \frac{1}{t-b} \, ,
\end{align}  
but also functions that depend on the elliptic curve $y$, such as
\begin{align} \label{eq:elliptic_kernel}
\Psi_{0}(b,t,\vec a) = \frac{c_4}{\omega_1 y} \, .
\end{align}  
For more details on the set of kernels that are allowed to appear, see~\cite{Broedel:2018qkq}. 

Individual $\mathcal{E}_{4}$ functions can be characterized by their length, which corresponds to the number of integrations $k$ in~\eqref{eq:cE4_def}, and their transcendental weight $\sum\lvert n_{i}\rvert$. Similar to multiple polylogarithms, they form a shuffle algebra, which allows one to write products of functions that share the same arguments $x$ and $\vec{a}$ as a sum of functions of higher length. They additionally obey identities such as the rescaling relation 
\begin{equation}
\cEfe{n_1 & \ldots & n_k}{p b_1 & \ldots& p b_k}{p x}{p \vec{a}} = \cEfe{n_1 & \ldots & n_k}{b_1 & \ldots& b_k}{x}{\vec{a}} \, ,
\end{equation}
which holds for any nonzero $p$ as long as $b_k \neq 0$. Finally, they have a symbol and coaction structure~\cite{2015arXiv151206410B}, which reduces to the normal coaction on multiple polylogarithms when one restricts to integration kernels such as~\eqref{eq:mpl_kernel} that don't depend on $y$. 

There are various benefits to defining elliptic multiple polylogarithms in the above way. The first is that integrals over kernels such as~\eqref{eq:elliptic_kernel} appear naturally in the integration of Feynman integrals, as seen for instance in the example of the two-loop banana in equation~\eqref{eq:two_loop_banana_integrated}. It is also possible to assign a notion of elliptic `purity' to the $\mathcal{E}_{4}$ functions~\cite{Broedel:2018qkq}, which extends the traditional notion of pure integrals as those whose (non-vanishing) maximal residues are all equal in magnitude~\cite{Arkani-Hamed:2010pyv}. For instance, expressed in terms of $\mathcal{E}_{4}$ functions, one finds that the two-loop banana integral is a pure function of weight two, the kite integral is a pure function of weight three, and the two-loop integrals in Figure~\ref{fig:ttbar} are pure functions of weight four~\cite{Broedel:2018qkq}. Finally, the $\mathcal{E}_{4}$ functions make manifest the Galois symmetry associated with the exchange of $y$ and $-y$, as these functions are either even or odd with respect to this exchange. 

Elliptic multiple polylogarithms can also be defined as iterated integrals on a complex torus, due to the map~\eqref{eq:elliptic_curve_torus_map}~\cite{BeilinsonLevin,LevinRacinet2007,brown2011multiple,Broedel:2017kkb,Broedel:2014vla,Matthes:2016}. The variant of these iterated integrals that is used most extensively in the evaluation of Feynman integrals is defined by\footnote{Another variant is defined by using a set of doubly periodic (but not meromorphic) function $f^{(n)}$ as the integration kernels, which is often used to evaluate string amplitudes~\cite{Broedel:2014vla}.}
\begin{equation} \label{eq:gamt_def}
    \gamt{n_1 & \ldots & n_k}{w_1 & \ldots& w_k}{z,\tau}=
    \int_{0}^{z}d z_{1}\,g^{(n_{1})}(z_{1}{-}w_{1},\tau)\gamt{n_{2} & \ldots & n_k}{w_{2} & \ldots& w_k}{z_{1},\tau} \, ,
\end{equation}
where the $n_i$ are integers, and the arguments $z$ and $z_i$ can be complex numbers. The integration kernel $g^{(n)}(z)$ are meromorphic quasi-doubly-periodic functions that can be generated by the Eisenstein-Kronecker series,
\begin{equation} \label{eq:gamt_kernel}
    \frac{\theta_{1}^\prime(0|\tau)\theta_{1}(z+\alpha|\tau)}{\theta_{1}(z|\tau)\theta_{1}(\alpha|\tau)} = \sum_{n\geq 0}\alpha^{n-1}g^{(n)}(z,\tau)\:. 
\end{equation}
Here $\theta_{1}$ is the odd Jacobi theta function, and $\theta_{1}^\prime$ is its derivative with respect to its first argument. One can easily translate between the $\mathcal{E}_4$ and $\tilde{\Gamma}$ functions, as the $\Psi_{n}$ kernels that appear in the former can be written as a linear combination of the kernels $g^{(|n|)}$ that appear in the latter~\cite{Broedel:2018qkq}.

A recursive formula exists for the total derivative of the $\tilde{\Gamma}$ functions, which makes them convenient for computing the symbol of an elliptic polylogarithm~\cite{Broedel:2018iwv}. As with multiple polylogarithms, the symbol can be used to derive identities between the $\tilde{\Gamma}$ functions. The total differential can also be used to express elliptic multiple polylogarithms that are evaluated at rational points $z = r/N + s \tau/N$ in terms of iterated Eisenstein integrals of level $N$, where $r$ and $s$ are integers~\cite{Manin,2014arXiv1407.5167B,Broedel:2018iwv}. This can be useful for evaluating these functions numerically, as the modular forms that appear as the kernels of these iterated Eisenstein integrals admit a Fourier expansion in the variable $q = e^{i \pi \tau}$, referred to as $q$-expansions. These expansions are guaranteed to converge since $\tau$ is always chosen to have a positive imaginary part. While this convergence can in general be slow, it can be sped up by finding a modular transformation that maximizes the imaginary part of $\tau$~\cite{Duhr:2019rrs,Weinzierl:2020fyx}. Algorithms for numerically evaluating elliptic multiple polylogarithms (including in terms of $q$-expansions) have already been implemented in {\sc GiNaC}~\cite{Bauer:2000cp,Walden:2020odh} (we will comment on them further in section~\ref{sec:numerics}).

While elliptic multiple polylogarithms are sufficiently general for evaluating many of the integrals discussed in section~\ref{sec:integral_zoo}, they can only be used in cases involving a single elliptic curve. In examples that involve multiple distinct elliptic curves, such as the double box in Figure~\ref{fig:ttbar_doublebox}, more general classes of iterated integrals are required~\cite{Chen}. However, this is not a fundamental limitation; in~\cite{Adams:2018bsn,Adams:2018kez}, this double box integral was evaluated to all orders in $\epsilon$ in terms of a space of iterated integrals involving 107 integration kernels and three separate elliptic curves (see also~\cite{Badger:2021owl}). It will be interesting to see whether a general class of functions for expressing Feynman integrals that involve multiple elliptic curves naturally suggests itself as more examples of this type are explored.

Certain Feynman integrals have also been found to be expressible in terms of more specialized spaces of functions. For instance, diagrams depending on just a single kinematic variable, such as the two- and three-loop banana diagrams with all equal internal masses and the three-loop contributions to the $\rho$ parameter, can be expressed in terms of iterated integrals of modular forms~\cite{Adams:2017ejb,Adams:2018ulb,Abreu:2019fgk,Broedel:2021zij}. The finite part of the two-loop banana integral with distinct masses has also been expressed in terms of functions that generalize the infinite sum representation of classical polylogarithms, namely
\begin{align} \label{eq:ELi_def}
\text{ELi}_{n;m}(x;y;q) = \sum_{j=1}^\infty \sum_{k=1}^\infty \frac{x^j}{j^n} \frac{y^k}{k^m} q^{jk} \, ,
\end{align}
where the argument $q$ is understood to be the nome $q = e^{i \pi \tau}$ of an elliptic curve~\cite{Adams:2014vja,Adams:2015gva}.\footnote{The negative of the nome, $-q$, also often appears as the third argument of the $\text{ELi}$ function; however, $\text{ELi}_{n;m}(x;y;-q)$ can always be rewritten as a linear combination of $\text{ELi}$ functions that instead depend on $q$.} 
Higher-depth versions of the $\text{ELi}$ function were also defined in~\cite{Adams:2015ydq,Adams:2016xah}, where they were used to evaluate the two-loop banana and the kite integral to all orders in the dimensional regularization parameter $\epsilon$. 

Finally, the two-loop banana graph with equal internal masses has also been evaluated in terms of the elliptic dilogarithm studied in~\cite{bloch2011higher}, which can be defined by
\begin{equation} \label{eq:elliptic_dilog}
D_\tau (\xi) = \sum_{n=-\infty}^\infty D\big(e^{2 \pi i \xi + 2 \pi i \tau n}\big) \, ,
\end{equation}
where $D(z)=\Im(\Li_{2}(z))+\arg(1-z)\log |z|$ is the Bloch-Wigner function. $D_\tau (\xi) $ is a particularly interesting function, as its study led to the discovery of the first non-trivial functional relations among elliptic polylogarithms~\cite{bloch2011higher,zagier2000classical,Broedel:2019tlz,2019arXiv190605068B}.
Such relations can be viewed as elliptic generalizations of the five-term identity for dilogarithms,
\begin{equation}
    D(x)+D(y)+D\biggl(\frac{1-x}{1-xy}\biggr)+D(1-xy)+D\biggl(\frac{1-y}{1-xy}\biggr)=0 \, ,
\end{equation}
which conjecturally generates all functional relations between dilogarithms. Conversely, it remains unclear whether other mechanism of generating functional relations between elliptic dilogarithms exist (although see~\cite{Remiddi:2017har,Broedel:2019tlz} for work on this topic). Techniques for reducing elliptic multiple polylogarithms to multiple polylogarithms (when possible) also remain relatively unexplored, although there exist examples in which Feynman integrals containing multiple non-rationalizable square roots have been found to be expressible in terms of multiple polylogarithms~\cite{Heller:2019gkq,Heller:2021gun,Kreer:2021sdt,Duhr:2021fhk}.

Importantly, methods for directly constructing the integrands of amplitudes using generalized unitarity~\cite{Bern:1994zx,Bern:1994cg,Britto:2004nc} have also recently been extended to the elliptic sector. Much of this work has been done in the context of `prescriptive' unitarity~\cite{Bourjaily:2013mma,Bourjaily:2015jna,Bourjaily:2017wjl,Bourjaily:2019iqr,Bourjaily:2019gqu,Bourjaily:2021vyj,Bourjaily:2021hcp,Bourjaily:2021ujs,Bourjaily:2021iyq}, in which one's basis of integrands is chosen so that each element directly matches a specific field theory cut. The coefficients of amplitudes in an appropriate prescriptive basis are then computable as `on-shell functions' as defined in \cite{Arkani-Hamed:2012zlh}. In~\cite{Bourjaily:2020hjv}, it was shown that the same approach can be extended to amplitudes involving elliptic curves, by generalizing the notion of leading singularities (traditionally used to refer to maximal-codimension residues) to any compact contour integral of maximal dimension. In particular, it was shown that on-shell functions defined for contours involving elliptic cycles enjoyed Yangian invariance in the case of planar maximally supersymmetric Yang-Mills theory. The role of elliptic leading singularities in prescriptive unitarity for the representation of these amplitudes was outlined in~\cite{Bourjaily:2021vyj}. 

The concepts of non-polylogarithmic leading singularities and prescriptive integrand bases allow us to probe the analytic structure of specific amplitudes in a surgical manner. For instance, to determine whether a particular amplitude is polylogarithmic or not, one can simply compute any non-polylogarithmic leading singularity of the amplitude; if it is non-vanishing, then the amplitude is non-polylogarithmic. Note that this follows from the mere possibility of constructing a prescriptive basis in which no other integrand has support on that integration contour, and does not require finding a prescriptive basis and using it to represent the amplitude. Moreover, these ideas are not limited to the elliptic case, but should generalize to arbitrary cases of higher rigidity if one considers integration cycles for instance over Calabi-Yau manifolds.\footnote{The rigidity of a Feynman integral is defined to be the smallest dimension over which it is non-polylogarithmic; see~\cite{Bourjaily:2018yfy}.}

\subsection{Integrals over Calabi-Yau Varieties}

The types of integrals over Calabi-Yau varieties that appear in Feynman integrals have been subject to much less study, both in the mathematics and the physics literature. In this section we briefly summarise what is known about them, mostly from the study of $L$-loop banana integrals.

Calabi-Yau varieties can be thought of as generalizations of elliptic curves to higher dimensions. More precisely, a Calabi-Yau $n$-fold is an $n$-dimensional complex variety $M_n$ with a unique nowhere-vanishing holomorphic $n$-form $\Omega_n$ (in other words, every holomorphic $n$-fold is a multiple of $\Omega_n$). Equivalently, one may say that the middle cohomology $H^{n,0}(M_n)$ of every Calabi-Yau manifold is one-dimensional. An elliptic curve is nothing but a Calabi-Yau one-fold, where the distinguished holomorphic one-form is the holomorphic differential $\Omega_1=dx/y$ on the elliptic curve.  

Typically, Feynman integrals give rise to families of Calabi-Yau $n$-folds parameterized by external kinematics. A basis of periods for these families of Calabi-Yau manifolds is obtained by integrating the holomorphic top-form $\Omega_n$ over a basis of the middle-dimensional homology group $H_n(M_n,\mathbb{Z})$. These periods, which are functions of the external kinematics, are the generalization of the periods of an elliptic curve from equations~\eqref{eq:period_1} and~\eqref{eq:period_2} to higher-dimensional Calabi-Yau varieties. Unlike in the elliptic case, the periods of Calabi-Yau $n$-folds for $n>1$ are in general not expressible in terms of known transcendental functions. However, several techniques have been developed in the context of geometry and string theory to find integral representations and/or locally converging power series for such periods; see for example~\cite{MR3965409}, and references therein. These periods (and their derivatives with respect to kinematic variables) also satisfy a set of quadratic relations, which are a direct consequence of the Hodge structure carried by the middle cohomology  $H^{n}(M_n)$. 

A special situation occurs for one-parameter families of Calabi-Yau varieties. In that case, the periods form a basis of solutions for the Picard-Fuchs differential operator attached to the Calabi-Yau variety, and these so-called Calabi-Yau operators and their solutions are expected to have particularly nice properties~\cite{MR3822913,BognerThesis,BognerCY}. In some instances, it is even possible to express the periods in terms of known hypergeometric functions. In particular, all periods of one-parameter families of Calabi-Yau two-folds can be written as products of elliptic integrals of the first kind~\cite{Doran:1998hm,BognerThesis,BognerCY}. 

Although periods of Calabi-Yau varieties have been studied in depth in the context of geometry and string theory, these periods are not the only integrals that arise in the context of Feynman integrals. Just like in the elliptic case, one also needs to consider integrals that involve (products of) periods in the integrand. For example, the $L$-loop equal-mass banana integral, considered in two space-time dimensions and with unit powers of the propagators, can be written as a one-fold integral of a period multiplied by a rational function. To the best of our knowledge, these types of integrals have never been considered in the literature, even in the one-parameter case. Developing a thorough understanding of this class of integrals would be interesting not only from a physics perspective, but also from the mathematics side, because they are the natural generalization of (iterated) integrals of modular forms to higher dimensional Calabi-Yau varieties. Currently, however, it is unclear what properties these integrals have.

\subsection{Numerical Evaluation}
\label{sec:numerics}

While expressing amplitudes in terms of well-studied classes of special functions is generally useful, being able to evaluate amplitudes numerically is sufficient for most phenomenological applications. As such, a variety of methods have been developed for numerically evaluating Feynman integrals, some of which can be applied to any process, and others of which are specialized to specific classes of integrals. We here review the most common methods, focusing on those that can be used to evaluate Feynman integrals that cannot be expressed in terms of multiple polylogarithms.

One of the more traditional approaches to numerically evaluating Feynman integrals is to use Monte Carlo (or quasi-Monte Carlo) integration techniques. Typically, these methods are applied to the Feynman-parameter representation given in~\eqref{eq:symanzik}. Since many Feynman integrals are divergent and need to be regulated, the method of \lq sector decomposition\rq ~\cite{Binoth:2000ps, Bogner:2007cr} is often applied to extract these singularities at the boundary of the integration region. Various programs have been developed over the years that implement sector decomposition~\cite{Bogner:2007cr, Smirnov:2008py, Carter:2010hi, Gluza:2010rn, Borowka:2017idc,Smirnov:2021rhf}, with \verb+FIESTA5+~\cite{Smirnov:2021rhf}, and \verb+pySecDec+~\cite{Borowka:2017idc} being the most recent and popular.

One of the chief advantages of sector decomposition is its broad applicability, which is not limited to any specific class of Feynman integrals. Unfortunately, this approach suffers from a few significant bottlenecks. First, a large number of points generally need to be evaluated to reach acceptable precision. Second, the computation of points in the physical region can require contour deformations that negatively affect the convergence of these methods. Finally, sector decomposition sometimes introduces a large number of terms in the integrand in order to resolve singularities.\footnote{We note, however, that this issue can be remediated by choosing a (quasi-)finite basis of Feynman integrals~\cite{Panzer:2014gra, vonManteuffel:2014qoa}.} As a result, although many impressive optimizations have been made in the latest implementations of these methods, they are typically less efficient and less precise than other methods, when others can be applied. 

More tailored techniques exist for specific use cases. For instance, a numerical integration strategy based on tropical geometry was introduced in~\cite{Borinsky:2020rqs}, which can be employed to evaluate Feynman integrals in the Euclidean region. In the framework of loop-tree duality~\cite{Catani:2008xa,Capatti:2019ypt, Capatti:2019edf, Capatti:2020ytd}, numerical integration methods have been developed that are suitable for the evaluation of finite Feynman integrals in integer numbers of dimensions. And algorithms and public codes~\cite{Czakon:2005rk, Gluza:2007rt, Gluza:2010rn, Dubovyk:2015yba, Usovitsch:2018shx} exist for evaluating Feynman integrals using Mellin-Barnes integral representations~\cite{Smirnov:1999gc, Tausk:1999vh}, which take the form of contour integrals that can be numerically integrated, or that can be computed as infinite sums over residues.

Another approach that has been pursued in recent years is to directly solve Feynman integrals numerically from the differential equations they satisfy, without making reference to an intermediate class of functions (see for example~\cite{Czakon:2013goa}). In particular, in~\cite{Mandal:2018cdj} it was shown that off-the-shelf numerical solvers can be used to solve differential equations of multi-loop Feynman integrals with good precision. A difficulty in this approach is that standard numerical solvers perform poorly near thresholds, where boundary conditions are easier to compute. Furthermore, non-trivial contour deformations are necessary to numerically cross thresholds singularities.

Alternatively, differential equations may be numerically solved in terms of one-dimensional generalized power series expansions along a line in the space of external momenta and internal masses (see for example~\cite{Pozzorini:2005ff,Aglietti:2007as,Mueller:2015lrx,Melnikov:2016qoc,Lee:2017qql,Melnikov:2017pgf,Liu:2017jxz,Lee:2018ojn,Bonciani:2018uvv,Mistlberger:2018etf,Bonciani:2018omm, Bruser:2018jnc,Davies:2018ood,Davies:2018qvx,Francesco:2019yqt, Frellesvig:2019byn, Bonciani:2019jyb, Abreu:2020jxa, Hidding:2020ytt,Liu:2020kpc,Abreu:2021smk,Fael:2021kyg,Becchetti:2021axs,Armadillo:2022bgm, Liu:2021wks,Liu:2022chg,Liu:2022tji}). These power series expansions are `generalized'  insofar as they may contain logarithms and algebraic roots that depend on the variable that parameterizes this line. By concatenating series solutions along multiple line segments, numerical solutions can be obtained in any region of phase-space. Contour deformations can be avoided by centering expansions at threshold singularities, and analytic continuation is only required for logarithms and algebraic roots that appear in the series solutions. A fully automated strategy was described in~\cite{Francesco:2019yqt}, and a public implementation was provided in the Mathematica package \verb+DiffExp+~\cite{Hidding:2020ytt}. The performance of series expansion methods greatly outperforms sector decomposition in most cases~\cite{Francesco:2019yqt}, and the method has a wide applicability (see for example~\cite{Pozzorini:2005ff,Caffo:2008aw,Abreu:2020cwb,
Badger:2021nhg,Dubovyk:2022frj,Bonciani:2021zzf,Abreu:2021smk,Badger:2021ega,Duhr:2021fhk,Fael:2021kyg,Abreu:2021asb,Chicherin:2021dyp,Becchetti:2021axs,Armadillo:2022bgm}). 

A remaining obstacle when solving Feynman integrals through differential equations, either numerically or analytically, is the determination of boundary conditions. Often this has been done analytically using expansion by regions~\cite{Beneke:1997zp, Smirnov:1999bza, Jantzen:2011nz}, or numerically using sector decomposition~\cite{Mandal:2018cdj}. The first approach is not completely automated, as it requires manual integration over various parametric integrals, whereas sector decomposition is automated but limited in precision. More recently, a Mathematica package \verb+AMFlow+ was released~\cite{Liu:2022tji}, which fully automates the determination of boundary conditions, and the subsequent numerical integration. More specifically, the package uses the auxiliary mass flow method~\cite{Liu:2017jxz, Liu:2020kpc,Liu:2021wks,Liu:2022chg,Liu:2022tji, Liu:2022mfb} to determine boundary conditions in a special limit of infinite complex mass. An internal solver based on series expansions, and optimized with a numerical fitting strategy, is then used to transport results to any region in phase-space. This allows the package to compute multi-loop integral families in a highly efficient manner, at high precision, and in a fully automated way.

Finally, in cases where a Feynman integral or amplitude is known in terms of certain classes of special functions, efficient numerical algorithms specific to these types of functions can be used. This has proven particularly useful for amplitudes that have been evaluated in terms of multiple polylogarithms, as these functions can be efficiently evaluated to high precision using algorithms implemented in {\sc GiNaC}~\cite{Bauer:2000cp,Vollinga:2004sn}. More recently, algorithms for numerically evaluating iterated integrals on $\mathcal{M}_{1,n}$ have also been implemented in {\sc GiNaC}~\cite{Walden:2020odh}. This class of functions includes elliptic multiple polylogarithms and iterated integrals of modular forms, and can thus be used to evaluate many of the integrals cataloged in section~\ref{sec:integral_zoo}. However, the code for evaluating iterated integrals on $\mathcal{M}_{1,n}$ still has some limitations. In particular, the analytic continuation of this class of functions is not yet automated, nor are identities for changing their arguments to speed up numerical convergence~\cite{Duhr:2019rrs,Weinzierl:2020fyx}. In part, this is because arbitrary modular transformations do not map these functions back to the same space of iterated integrals; however, it has been shown that this problem can be overcome by combining modular transformations with a change of basis of master integrals~\cite{Weinzierl:2020fyx}. At the moment, this combined transformation must be carried out by hand. 

While the classes of special functions required for evaluating scattering amplitudes that involve integrals over Calabi-Yau manifolds are still under study, numerical results for the equal-mass banana integrals have recently been computed through four loops~\cite{Bonisch:2021yfw}. This was done by constructing a set of generalized power series solutions to the Picard-Fuchs operator that describes the Calabi-Yau geometry of these integrals (derived using the Bessel function representation of these integrals~\cite{Berends:1993ee,Bauberger:1994by,Bauberger:1994hx,Bauberger:1994nk,Vanhove:2014wqa}), using the Frobenius method as described in~\cite{Bonisch:2020qmm} and implemented there in {\sc PARI/GP}~\cite{PARI2}. These generalized power series can be efficiently evaluated to get the numerical values of these integrals. The same methods can in principle also be used to numerically evaluate the banana integral with generic masses, but this requires constructing a generalized power series in many variables at once, which is significantly more onerous in practice.


\section{Open Questions and Directions for Future Research}
\label{sec:future_research}

Despite the substantial progress covered in the last two sections, there remain many open questions about Feynman integrals that involve integrals over elliptic curves and Calabi-Yau manifolds. Below, we highlight some of the research directions that constitute important avenues for future work. 
\\[-.3cm]

\noindent {\bf Phenomenological Calculations Involving Many Kinematic Variables --} A number of new multi-loop calculations will be required in the coming years to achieve the theoretical precision demanded by future collider programs. In general, the Feynman integrals entering these calculations will depend on increasing numbers of kinematic variables, as well as large numbers of massive propagators. Correspondingly, the analytic structure of these integrals is also expected to increase in complexity. This makes semi-numerical approaches to evaluating these integrals (for instance, using generalized series expansions~\cite{Mandal:2018cdj, Francesco:2019yqt, Hidding:2020ytt, Liu:2022chg}) especially attractive. Nevertheless, in order to make use of these numerical techniques, it will be important to construct bases of master integrals that satisfy differential equations that are linear in the dimensional regularization parameter. As this generally becomes more difficult as the number of kinematic variables increases (due to the appearance of more square roots, or integrals over multiple algebraic varieties), a great deal of future work will be devoted to the construction of such bases for phenomenological applications, for example in five-point processes such as $t \bar{t}+$jet production and $t \bar{t} H$ production~\cite{Amoroso:2020lgh,Chicherin:2021dyp,Badger:2022mrb}.
 \\[-.3cm]
 
 \noindent {\bf Epsilon Factorization and Generalized Canonical Forms --} For Feynman integrals in dimensional regularization, the method of differential equations combined with the idea of canonical forms has been invaluable. How much of this approach can be generalized beyond the polylogarithmic case? There are several examples in which epsilon-factorized differential equations have been obtained for elliptic Feynman integrals~\cite{Adams:2018yfj, Bogner:2019lfa}, and it is known how to generate such forms algorithmically~\cite{Frellesvig:2021hkr}. Yet integrating these systems in terms of special functions such as elliptic multiple polylogarithms is not nearly as easy as in the case of multiple polylogarithms. For multiple polylogarithms, the key property that made integration simple was the fact that everything was already expressed in terms of $d \log$ forms. Finding a suitable generalization of this property to the elliptic case (and beyond) that similarly allows for straightforward integration would greatly benefit future calculations.
 \\[-.3cm]

\noindent {\bf Development of Future Numerical Tools --} A number of algorithmic and implementational improvements can be made in the numerical solvers currently being used to evaluate Feynman integrals. For example, recent progress was made in~\cite{Liu:2022chg}, where a fast and parallelizable solver was developed by numerically fitting the dimensional regulator. Speedups may also be attainable by porting current codes written in Mathematica to lower-level languages such as C++. Lastly, the construction of differential equations requires solving complicated systems of IBP relations, which in many cases form a computational bottleneck; for this reason, further optimizations on the side of IBP codes are likely to be important in the future.
\\[-.3cm]

\noindent {\bf Symbolic Computational Tools for Elliptic Multiple Polylogarithms --} Although public codes for integrating and manipulating multiple polylogarithms have been available for a number of years~\cite{Maitre:2005uu,Maitre:2007kp,Panzer:2014caa,Duhr:2019tlz}, there is currently no public code for working with elliptic multiple polylogarithms. The development of publicly-available software for directly integrating elliptic multiple polylogarithms, computing their symbol, and converting between the different formulations of these functions on elliptic curves and complex tori would be of great benefit to the community.
\\[-.3cm]
 
 \noindent {\bf Identities Between Iterated Integrals --} Iterated integrals beyond elliptic multiple polylogarithms are expected to be required in the evaluation of many phenomenologically-relevant scattering amplitudes. In such cases, it will prove useful (or even essential) to be able to reduce to a basis of functions by finding algebraic relations between these iterated integrals. For instance, a space of iterated integrals involving three distinct elliptic curves naturally arises in the calculation of $t \bar{t}$ production via gluon fusion, due to the contribution from the double box integral in Figure~\ref{fig:ttbar_doublebox}~\cite{Adams:2018bsn,Adams:2018kez}. It was found that nontrivial identities between these iterated integrals were required to make the cancellation of ultraviolet and infrared poles manifest in the two-loop QCD amplitudes for this process~\cite{Badger:2021owl,Chaubey:2021ret}. Such relations also tend to lead to massive simplifications in the functional form of scattering amplitudes, making them easier to work with and evaluate numerically. It will correspondingly be important to develop general methods for finding such identities in phenomenological applications in the future. Notably, one approach for doing so was developed in~\cite{Remiddi:2016gno,Remiddi:2017har}, where it was shown that functional relations between iterated integrals can be derived using higher-order differential operators that annihilate the integration kernels appearing in these functions.
\\[-.3cm]

\noindent {\bf Cataloging Integration Geometries --} Due to the existence of integrand-level basis reduction techniques (such as generalized unitarity and its variants~\cite{Melrose:1965kb,Passarino:1978jh,Bern:1994zx,Bern:1994cg,Britto:2004nc,Ossola:2006us,Mastrolia:2010nb,Bourjaily:2017wjl,Bourjaily:2013mma,Bourjaily:2015jna,Bourjaily:2020qca}) and IBP identities~\cite{Chetyrkin:1981qh,Laporta:2000dsw}, a comparatively small number of independent Feynman integrals contribute to scattering amplitudes at low loop orders. This should make it possible to catalog all the algebraic varieties that appear in these integrals at a given loop order, especially in theories of interest. Such a catalog would provide valuable information about the class of special functions required for expressing amplitudes in different theories and different loop orders. Moreover, it will help determine whether there exist any universal features of the geometries that arise in Feynman integrals, and whether these features can be tied to basic physical principles.
\\[-.3cm]

\noindent {\bf Uniqueness of Geometries --} An important open question is whether the geometries associated with a Feynman integral are unique. In the elliptic case, the two-loop sunrise integral can be described in terms of two distinct, but isogenous, elliptic curves~\cite{Adams:2017ejb,Bogner:2019lfa}. While both curves can be useful for representing the integral in different contexts, it was found in~\cite{Frellesvig:2021vdl} that only one of the curves corresponds directly to the geometry of the initial integrand, while the other curve was introduced by changes of variables involving a double-cover. Beyond the elliptic case, some of the authors at one point speculated that every Feynman integral may be associated with a Calabi-Yau geometry. This cannot be true at face value, as there are Feynman integrals associated with higher-genus Riemann surfaces, which are not Calabi-Yau~\cite{Huang:2013kh,Hauenstein:2014mda}. However, the approach taken in~\cite{Bonisch:2021yfw} suggests that these integrals may still be associated with Calabi-Yau motives, in which distinct geometries are identified when they share a particular subspace of their cohomology. This opens up the possibility that each Feynman integral could be associated to a unique Calabi-Yau motive, which would be a powerful tool in investigating and classifying these integrals. In particular, if true, this might suggest an approach to evaluating diagrams involving multiple elliptic curves~\cite{Adams:2018bsn,Adams:2018kez} in which these curves were combined into a single Calabi-Yau motive.
\\[-.3cm]

\noindent {\bf Non-Polylogarithmic Leading Singularities Beyond Elliptic Curves --} Although the definition of leading singularities introduced in~\cite{Bourjaily:2020hjv} naturally extends to cases involving an arbitrary compact contour integral, those involving geometries that go beyond elliptic curves are stymied by a lack of clear computational methods for parameterizing such contours, or the existence of standard functions in terms of which these integrals can be expressed. It will be important to develop the technology for computing these higher-rigidity leading singularities over the coming years. 
\\[-.3cm]

\noindent {\bf Special Functions for Integrals over Calabi-Yau Manifolds --} One of the most important open research directions concerns the development of special functions for expressing scattering amplitudes that involve Calabi-Yau manifolds. The first step towards understanding this space of functions would involve classifying the independent integration kernels that can appear in Feynman integrals of this type, and how these kernels depend on the details of the Calabi-Yau geometry. Since there is always some freedom in the choice of an independent set of integration kernels, an important consideration will be whether these kernels can be chosen such that physical principles such as locality can be made manifest. Another basic requirement should be that the kernels only involve logarithmic singularities. Once a basis of kernels has been selected, one would then have to work out the basic technology for working with this space of iterated integrals, including techniques for analytically continuing them and for generating sum representations that can be efficiently evaluated numerically. For an example of how such technology can be developed, see recent work on iterated Eisenstein integrals~\cite{Manin,2014arXiv1407.5167B,Adams:2017ejb,Broedel:2018iwv,Duhr:2019rrs,Broedel:2019kmn}.
\\[-.3cm]

\noindent {\bf Hyperelliptic Integrals --} In addition to elliptic curves and Calabi-Yau manifolds, integrals over hyperelliptic curves---or, curves of genus greater than one---have been observed in Feynman integrals~\cite{Huang:2013kh,Hauenstein:2014mda}. These types of integrals also appear in open-string amplitudes starting at two loops, and thus can be studied in the clean laboratory provided by string theory. For each surface that appears in the genus expansion of open-string amplitudes, one should be able to find a set of holomorphic differentials that respect the symmetries of the surface (by being at least quasi-periodic with respect to all its cycles), and a canonical connection form analogous to the Eisenstein-Kronecker series~\eqref{eq:gamt_kernel} that appears at genus one. For these higher-genus surfaces, the modular parameter $\tau$ will be replaced by a Riemann period matrix in the Siegel upper half plane, Jacobi $\vartheta$-functions will be replaced by $\vartheta$-functions with character, and $\mathrm{SL}(2,\mathbb{Z})$ invariance will become $\mathrm{Sp}(2n,\mathbb{Z})$ invariance. One can also consider the function spaces that one encounters by integrating along surfaces with boundary, which should allow one to find relations between higher polylogarithmic functions and their single-valued counterparts.   
\\[-.3cm]

\noindent {\bf Bootstrapping Elliptic Amplitudes --} Efficient methods for bootstrapping certain polylogarithmic amplitudes, form factors, and anomalous dimensions have been developed in recent years (see for instance~\cite{Caron-Huot:2020bkp,Li:2016ctv,Almelid:2017qju,Dixon:2020bbt}). These methods usually make use of the coproduct in order to iteratively construct the space of functions within which these quantities are expected to exist, after which physical constraints are imposed until a unique function is found. The same type of coproduct structure exists for elliptic polylogarithms~\cite{Broedel:2018iwv,2015arXiv151206410B}, and it would be interesting to use this structure to bootstrap elliptic amplitudes or Feynman integrals in the same way. Key ingredients for such a program would be a good understanding of the space of elliptic (and non-elliptic) symbol letters that appear in the quantity under study, as well as a set of symmetries and adequate boundary data with which to constrain an ansatz within this space.
\\[-.3cm]

\noindent {\bf Elliptic Coaction Principles --} A number of amplitudes and Feynman integrals have been observed to exhibit interesting number-theoretic symmetries associated with the so-called cosmic Galois group~\cite{Cartier2001,2008arXiv0805.2568A,2008arXiv0805.2569A,Brown:2015fyf}. Namely, these quantities exhibit a certain type of stability under the coaction, insofar as the functions that appear in the first entry of their coaction correspond to functions that appear in the same quantity at lower loop orders. This type of `coaction principle' was first studied in massless $\phi^4$ theory~\cite{Schnetz:2013hqa,Brown:2015fyf,Panzer:2016snt}, and has since been observed in the polylogarithmic part of the anomalous magnetic moment of the electron~\cite{Laporta:2017okg,Schnetz:2017bko}, in amplitudes and form factors in maximally supersymmetric gauge theory~\cite{Caron-Huot:2019bsq,Dixon:2020bbt}, and in integrable fishnet models~\cite{Gurdogan:2020ppd}. Similar structures have also been observed in string perturbation theory, where different orders in the $\alpha'$ expansion of tree-level string amplitudes are stable under the coaction rather than different loop orders~\cite{Schlotterer:2012ny}. It will be interesting to see if such coaction principles also exist beyond the multiple polylogarithmic sector.
\\[-.3cm]
 
\noindent {\bf Isolating Integral Contributions over Specific Manifolds --} In~\cite{Bourjaily:2019iqr,Bourjaily:2019gqu,Bourjaily:2021hcp,Bourjaily:2021iyq}, a set of Feynman integral numerators was found that made the cancellation of all non-polylogarithmic contributions to the two-loop MHV amplitudes in maximally supersymmetric gauge theory manifest at the level of the integrand, even though elliptic and K3 contributions show up in individual integrals. It would be interesting to know to what extent the contribution from specific integration contours can be isolated in more general bases of Feynman integrals using similar techniques. 
\\[-.3cm]

\noindent {\bf Stratifications of Rigidity --} In \cite{Bourjaily:2021vyj} it was shown that most dual-conformal integrands at two loops must involve both polylogarithmic and non-polylogarithmic parts. Moreover, many such integrals involve multiple elliptic curves. It is thus natural to wonder whether these features can be avoided. There is now considerable evidence that the answer is yes, at least for the case of planar amplitudes at two loops. Specifically, by abandoning dual-conformal invariance (by increasing the power-counting of the basis), it is possible to construct a basis which is fully divided into polylogarithmic and non-polylogarithmic sub-strata, and in which no Feynman integral involves more than one elliptic curve~\cite{stratifyingRigidity}. It remains unclear the extent to which these features may be generalized to higher-loop orders, or to non-planar integrand bases at two loops (as either case would involve integrals with higher rigidity). 
\\[-.3cm]

\noindent {\bf Degenerations of Integral Geometries and Landau Singularities --} As mentioned in section~\ref{sec:review_and_importance}, the varieties that appear in Feynman integrals often degenerate in special kinematic limits. For the two-loop banana integral discussed in section~\ref{sec:banana-integrals} this happens at thresholds and pseudo-thresholds corresponding to solutions of the Landau equations (see for example~\cite{Frellesvig:2021vdl}). Making this connection more precise for more complicated integrals with more kinematic parameters could provide interesting insight into the role played by the moduli spaces of the geometries that appear in these integrals. A natural starting point are the traintrack integrals in section~\ref{sec:traintrack}, whose geometry has been described in momentum twistor space~\cite{Caron-Huot:2012awx,Vergu:2020uur}.
\\[-.3cm]

\noindent {\bf Landau Analysis and the Coaction --} Important work on how the Landau equations~\cite{landau1959} and cut integrals~\cite{Cutkosky:1960sp} can be used to constrain the analytic structure of Feynman integrals (and in some cases even reconstruct the full motivic coaction of these integrals) has been carried out in the context of multiple polylogarithms~\cite{Abreu:2014cla,Dennen:2015bet,Bloch:2015efx,Kreimer:2016tqq,Dennen:2016mdk,Prlina:2017azl,Abreu:2017ptx,Prlina:2017tvx,Prlina:2018ukf,Bourjaily:2020wvq,Benincasa:2020aoj,Hannesdottir:2021kpd}. Moreover, it has been shown that Feynman integrals at low loop orders are endowed with a diagrammatic coaction that agrees with the motivic coaction on multiple polylogarithms at all orders in the $\epsilon$ expansion in dimensional regularization~\cite{Abreu:2017enx,Abreu:2017mtm,Abreu:2018sat,Abreu:2018nzy,Abreu:2019eyg,Brown:2019jng,Abreu:2021vhb,Gardi:2021qov}. It would be interesting to see how these analyses generalize to cases involving iterated integrals over elliptic curves and the periods of higher-dimensional manifolds.
\\[-.3cm]

\noindent {\bf Singularities and Cluster Algebras --} A number of surprising connections have been observed between the singularity structure of scattering amplitudes (or even individual Feynman integrals) and cluster algebras (or algebraic and geometric structures closely related to cluster algebras, including tropical fans and polytopes)~\cite{Arkani-Hamed:2012zlh,Golden:2013xva,Golden:2014xqa,Golden:2014pua,Drummond:2017ssj,Drummond:2018dfd,Golden:2018gtk,Drummond:2018caf,Golden:2019kks,Drummond:2019qjk,Drummond:2019cxm,Arkani-Hamed:2019rds,Henke:2019hve,Drummond:2020kqg,Mago:2020kmp,Chicherin:2020umh,Mago:2020nuv,Herderschee:2021dez,He:2021esx,Mago:2021luw,Henke:2021ity,Ren:2021ztg}. However, all of these connections currently remain restricted to amplitudes that can be expressed in terms of multiple polylogarithms. It would be interesting to explore whether the elliptic and Calabi-Yau varieties that appear at higher loop order and particle multiplicity are also encoded in suitable generalizations of these algebraic or geometric structures.
\\[-.3cm]

\noindent {\bf Cycle of Calculation and Discovery --} It should not be overlooked that progress is made towards the research goals suggested above every time new results for previously-unknown integrals become available. Indeed, the history of progress in the amplitudes field suggests that this field is largely data-driven. Great effort is often expended to evaluate some new integral that is barely possible with existing technology. Over time, as a critical mass of data accumulates in the literature, it becomes possible to observe, extrapolate, or merely guess properties that enable the next round of previously-impossible calculations. Therefore, in addition to the lofty goals elaborated above, no effort should be spared on attacking all varieties of new integrals, if for no other purpose than to fuel this virtuous feedback loop.
\\[-.3cm]

\section{Outlook}
\label{sec:conclusions}

Scattering amplitudes often involve integrals over complicated algebraic varieties such as elliptic curves and Calabi-Yau manifolds. These types of integrals start to appear in precision QCD calculations at NNLO and in supersymmetric gauge theory at two loops, making them of high experimental and theoretical interest. Thus, while substantial progress has been made in the development of technology for evaluating Feynman integrals that involve elliptic curves, it will be important to continue to expand our analytic and numerical control over these types of integrals, and to develop similar technology for evaluating integrals over higher-dimensional varieties. In this white paper, we have highlighted a number of specific research directions in which impactful progress can be made towards these goals. Given the important role that precision standard model predictions will play in the search for new physics in future colliders, and the insight this work will give us into the mathematical structure of perturbative quantum field theory, these topics deserve to be a focal point of research in the coming years.

\acknowledgments

JLB is supported by a grant from the US Department of Energy (DE-SC00019066), and JLB, RM, CV, MvH, MW, and CZ are supported by a Starting Grant (No. 757978) from the European Research Council. EC receives funding from the European Union's Horizon 2020 research and innovation programmes, New level of theoretical precision for LHC Run 2 and beyond (grant agreement No.~683211), and High precision multi-jet dynamics at the LHC (grant agreement No.~772009). HF has received funding from the European Union's Horizon 2020 research and innovation program under the Marie Sk{\l}odowska-Curie grant agreement No.~847523 `INTERACTIONS', as well as a Carlsberg Foundation Reintegration Fellowship. MH is supported by the European Research Council under ERC-STG-804286 UNISCAMP. RM, MW, and CZ are additionally supported by the research grant 00025445 from Villum Fonden. MS and AV are supported by the US Department of Energy under contract DE-SC0010010 Task A, and AV is additionally supported by Simons Investigator Award \#376208. LT was supported by the Excellence Cluster ORIGINS funded by the Deutsche Forschungsgemeinschaft (DFG, German Research Foundation) under Germany's Excellence Strategy - EXC-2094 - 390783311 and by the ERC Starting Grant 949279 HighPHun.


\bibliographystyle{JHEP}
\bibliography{beyond_polylogs}

\end{fmffile}
\end{document}